\documentclass{aa}  

\usepackage{graphicx}
\usepackage{txfonts}
\usepackage{hyperref}
\usepackage{array}
\usepackage{caption}
\usepackage{makecell}
\usepackage{multirow}
\usepackage{xcolor}
\newcommand{\para}[1]{\left(#1\right)}
\newcommand{\cro}[1]{\left[#1\right]}

\newcommand{\abs}[1]{\left\vert #1  \right\vert}
\newcommand{\rz}{r_0}

\newcommand{\otf}[1]{\tilde{h}_{#1}}

\newcommand{\psd}[1]{\mathcal{W}_{#1}}
\newcommand{\cov}[1]{\mathcal{B}_{#1}}
\newcommand{\dphi}[1]{\mathcal{D}_{#1}}
\newcommand{\rhol}{\boldsymbol{\rho}/\lambda}
\newcommand{\rhovec}{\boldsymbol{\rho}}
\newcommand{\rvec}{\boldsymbol{r}}
\newcommand{\ccc}[1]{\textcolor{black}{#1}}
\newcommand{\pup}{\mathcal{P}}
\newcolumntype{P}[1]{>{\centering\arraybackslash}p{#1}}

\begin{document}

  %\title{Exhaustive demonstration of an analytical point spread function model for adaptive-optics assisted optical/near infra-red instruments.  }
  
  \title{Joint estimation of atmospheric and instrumental defects using a parsimonious point spread function model.}
   \subtitle{On-sky validation using state of the art worldwide adaptive-optics assisted instruments.}
\titlerunning{Joint estimation of atmospheric and instrumental defects}
\authorrunning{O. Beltramo-Martin et. al.}

\author{Olivier Beltramo-Martin\inst{1,2}, Romain F{\'e}tick\inst{2,1},  Benoit Neichel\inst{1},Thierry Fusco\inst{2,1}}

\institute{ 
 Aix Marseille Univ, CNRS, CNES, LAM, Marseille, France\ \and
DOTA, ONERA, Université Paris Saclay, F-91123 Palaiseau, France\
 }
   \date{}

% \abstract{}{}{}{}{} 
% 5 {} token are mandatory
 
\abstract
{Modeling the optical point spread function (PSF) is particularly challenging for adaptive optics (AO)-assisted observations owing to the its complex shape and spatial variations.}
{We aim to (i) exhaustively demonstrate  the accuracy of a recent analytical model from comparison with a large sample of imaged PSFs, (ii) assess the conditions for which the model is optimal, and (iii) unleash the strength of this framework to enable the joint estimation of atmospheric parameters, AO performance and static aberrations.}
{We gathered 4812 on-sky PSFs obtained from seven AO systems and used the same fitting algorithm to test the model on various AO PSFs and diagnose AO performance from the model outputs. Finally, we highlight how this framework enables the characterization of the so-called low wind effect on the Spectro-Polarimetic High contrast imager for Exoplanets REsearch (LWE ; SPHERE) instrument and piston cophasing errors on the Keck II telescope.}
{ Over 4812 PSFs, the model reaches down to 4\% of error on both the Strehl-ratio (SR) and full width at half maximum (FWHM). We particularly illustrate that the estimation of the Fried's parameter, which is one of the model parameters, is consistent with known seeing statistics and follows expected trends in wavelength using the Multi Unit Spectroscopic Explorer (MUSE) instrument ($\lambda^{6/5}$) and field (no variations) from \ccc{Gemini South Adaptive Optics Imager (GSAOI)} images with a standard deviation of 0.4\,cm.  Finally, we show that we can retrieve a combination of differential piston, tip, and tilt modes introduced by the LWE that compares to ZELDA measurements, as well as segment piston errors from the Keck II telescope and particularly the stair mode that has already been revealed from previous studies. }
{This model matches all types of AO PSFs at the level of 4\% error and can be used for AO diagnosis, post-processing, and wavefront sensing purposes. } 


\maketitle

\section{Introduction}
Adaptive optics (AO, \citealt{Roddier1999}) is a game changer in the quest for high-angular resolution, especially for ground-based astronomical observations that face the presence of wavefront aberrations introduced by the atmosphere \citep{Roddier1981}. Thanks to AO, the point spread function (PSF) delivered by an optical instrument is much narrower, by a factor up to \ccc{50} on the full width at half maximum (FWHM), than the seeing-limited scenario \citep{Roddier1981}. Still, some correction residuals persist and render the AO PSF shape complex to model. Consequently, standard parametric models that reliably reproduce the seeing-limited PSFs, such as a Moffat function \citep{Trujillo2001,Moffat1969}, become inefficient at describing the AO PSF. Moreover, contrary to seeing-limited observations, AO-corrected images suffer from the anisoplanatism effect \citep{Fried1982} that strengthens the spatial variations of the PSF on top of instrument defects. Determining the AO PSF is necessary for two major reasons. \ccc{Firstly,
understanding and accurately modeling the PSF morphology is key to diagnosing AO performance.} 
From the PSF, we can identify the major contributors to the AO residual error \citep{BeltramoMartin2019_PRIME,Ferreira2018,Martin2017}. Secondly, the image delivered by an optical instrument depends on both the science object we want to characterize and the PSF. In order to estimate the interesting astrophysical quantities, one can either use a deconvolution technique \citep{Fetick2020,Fetick2019_VESTA,Benfenati2016,Flicker2005,Mugnier2004,Fusco2003,Drummond1998} or include the PSF as part of a model, as is performed in PSF-fitting astrometry/photometry retrieval techniques \citep{BeltramoMartin2019_PRIME,Witzel2016,Schreiber2012,Diolaiti2000,Bertin1996,Stetson1987} and galaxy kinematics estimation \citep{Puech2018,Bouche2015,Epinat2010} for instance.

Nevertheless, the PSF is not systematically straightforward to determine from the focal-plane image, particularly for observations made with a small field of view (FOV), or of extended objects, crowded populations or with a low signal-to-noise ratio (S/N). Alternative methods exist, such as PSF reconstruction \citep{BeltramoMartin2020_ZIMPOL_PSFR,BeltramoMartin2019_PRIME,Wagner2018,Gilles2018,Ragland2018_PSFR,Martin2016JATIS,Ragland2016,Jolissaint2015,Exposito2014,Clenet2008,Gendron2006,Veran1997}, which is potentially very accurate (1\% error level). However, this process lacks science verification and requires years to be fully implemented and operational as a stand-alone pipeline. The inherent complexity has inspired several efforts to produce simpler but still efficient reconstruction techniques \citep{Fusco2020,Fetick2019_Moffat_aa} that have been fully validated and are now in operation for Multi Unit Spectroscopic Explorer (MUSE) \citep{Bacon2010} at the Very Large Telescope (VLT).
The analytic AO PSF model described by \citet{Fetick2019_Moffat_aa} has shown spectacular accuracy in reproducing the  MUSE Narrow Field Mode (NFM) PSF as well as the PSF of the Zurich Imaging Polarimeter (ZIMPOL) \citep{Schmid2018} that equips the Spectro-Polarimetic High contrast imager for Exoplanets REsearch (SPHERE) instrument \citep{Beuzit2019} PSF and is an excellent candidate to rethink the way we perform PSF reconstruction, with a single flexible algorithm that complies with all kinds of AO correction. However, several questions arise as to its capabilities. Does this model really match any type of AO PSF, regardless of the observing conditions and even at very high Strehl-ratio (SR) ? How much accuracy can we expect in a statistical sense and what are the fundamental limits ? Can we improve the description of the PSF in the presence of static aberrations ?
The goal of this paper is to address these questions by (i) providing an exhaustive demonstration that this model complies with any type of AO system, telescope, and in optical and near infrared (NIR) wavelengths, (ii) assessing conditions for which the model is optimal for representing an AO PSF, and (iii) showing that this framework is robust enough to enable the joint estimation of AO residual and instrumental aberrations. In Sect. \ref{S:th}, we present the theoretical background and the model description as well as some upgrades we have implemented so as to treat new types of data and include static-aberration fitting. In Sect. \ref{S:stat}, we present a statistical analysis of the model accuracy over 4812 PSFs with a discussion on the parameter estimates. 
Finally, in Sect. \ref{S:static},  we illustrate how the present model allows  joint estimation of atmospheric conditions, AO performance and static aberrations, caused by the low wind effect (LWE) on SPHERE or segment cophasing error at Keck.
\section{PSF model}
\label{S:th}
This model was originally introduced by \citet{Fetick2019_Moffat_aa} and validated on SPHERE/ZIMPOL and MUSE NFM data. In order to increase the range of applicability of such a model, we upgraded it by including (i) \ccc{the model of the Apodized Lyot Coronagraph (APLC, \citet{Soummer2005}) used on SPHERE (focal-plane mask not considered) }
 (ii) additional degrees of freedom to adjust static aberrations over a specific modal basis, which can be Zernike modes described on the pupil, tip-tilt and piston modes for each pupil area delimited by the spiders so as to describe the so-called LWE or piston modes for each pupil segment \citep{Laginja2019, Leboulleux2018, Ragland2018_COPHASING}.

First of all, the model assumes the stationarity of the phase of the electric field in the pupil plane \citep{Roddier1981}, which allows the system optical transfer function (OTF) to be split into a static part $\otf{\text{Static}}$ (telescope + internal aberrations) and an AO residual spatial filter $\tilde{k}_\text{AO}$ as follows,
\begin{equation} \label{E:otfdot}
    \otf{}(\rhol) = \otf{\text{Static}}(\rhol) \tilde{k}_\text{AO}(\rhol),
\end{equation}
where $\rho$ is the separation vector within the pupil, $\lambda$ is the imaging wavelength, and $\rhol$ is the angular frequencies vector.

\begin{figure}[h!]
    \centering
    \includegraphics[width = 0.49\columnwidth]{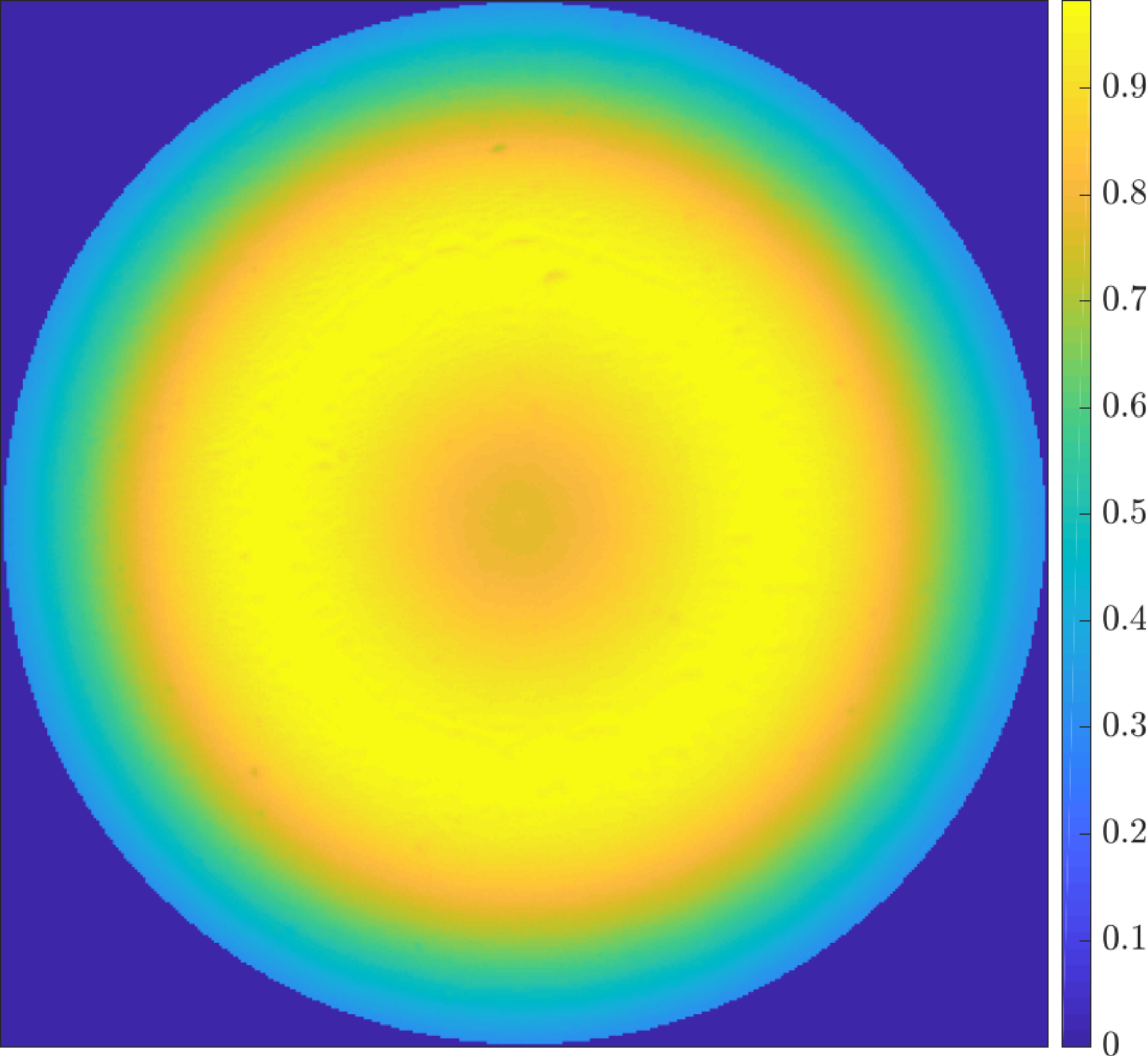}
    \hspace{0.02\columnwidth}
    \includegraphics[width = 0.45\columnwidth]{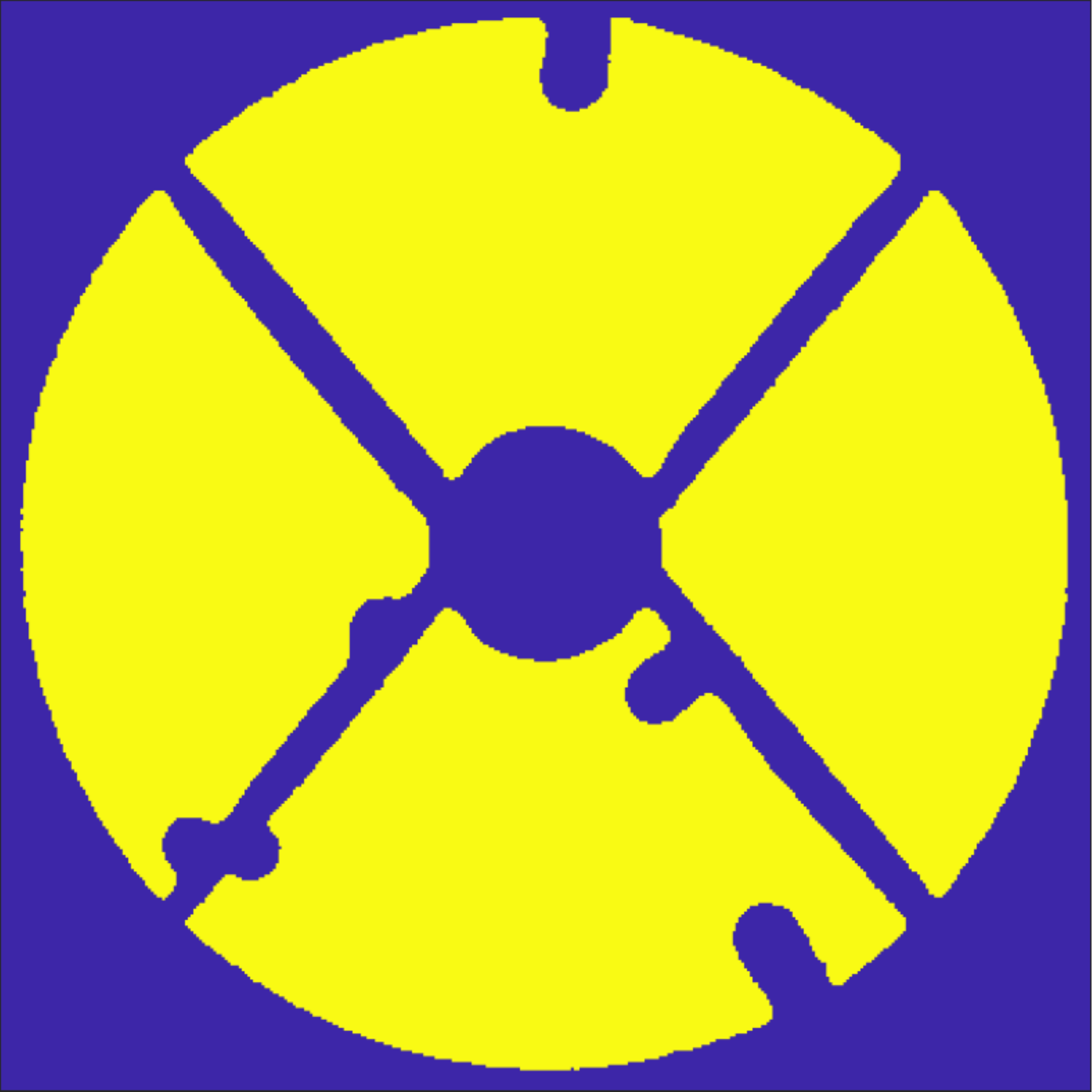}
    \caption{From left to right: Pupil plane apodizer and Lyot stop used during SPHERE/IRDIS observations with the N\_ALC\_YJH\_S APLC  \citep{Vigan2010MNRAS} .}
    \label{fig:static}
\end{figure}

The static OTF derives from the instrument pupil mask $\pup(\rvec)$ and the optical path difference (OPD) map $\Delta_\text{Static}$ in the pupil plane, which gives for a monochromatic beam at wavelength $\lambda$
\begin{equation} \label{E:otfStatic}
\begin{aligned}
    &\otf{\text{Static}}(\rhol)  =\\ &\iint_\pup \pup(\rvec)\pup^*(\rvec+\rhovec)\exp\para{2i\pi/\lambda(\Delta_\text{Static}(\rvec) - \Delta_\text{Static}(\rvec+\rhovec) )}\boldsymbol{dr}.
\end{aligned}
\end{equation}
The SPHERE/IRDIS data we have treated (see Sect. \ref{SS:data} for a more complete description) were obtained during PSF calibration procedure using an off-axis stars. Therefore, the incoming beam was not getting through the focal-plane mask and the pupil mask model results from the multiplication of the apodizer (amplitude) function $\mathcal{A}$ and the Lyot stop $\mathcal{L}$ \citep{Soummer2005} as follows
\begin{equation} \label{E:apodizer}
    \pup(\rvec) = \mathcal{A}(\rvec).\mathcal{L}(\rvec).
\end{equation}
The 2D functions $\mathcal{A}(\rvec)$ and $\mathcal{L}(\rvec)$ are illustrated in Fig. \ref{fig:static}.

In this paper, we split the  static aberration contribution in two terms: the  Non Common Path Aberrations (NCPA) calibration (non-coronagraphic) noted $\Delta_\text{NCPA}$  \citep{Jia2020,Vigan2019,Lamb2018,Vassallo2018,Ndiaye2016,Sitarski2014,Jolissaint2012,Sauvage2011,Clelia2008,Mugnier2008,Blanc2003,Fusco2003}, to which is added another contribution decomposed over a modal basis $\boldsymbol{M}$ to be specified. This latter could be simply Zernike modes over the whole pupil or per pupil area delimited by the spiders, so as to include potential LWE that may introduce severe asymmetries in the PSF structure. Analyses of SPHERE images \citep{Milli2018_LWE,Sauvage2016,Sauvage2015} showed that the LWE mainly introduces piston, tip, and tilt differential aberrations between pupil areas separated by the spiders, which represents 12 parameters to be adjusted. Also, for segmented pupil telescopes like Keck telescopes, this basis can be segment piston or petal modes \citep{Ragland2018_COPHASING}.  We analyze the capacity of this model to identify such aberrations in Sect. \ref{S:static}. The static OPD is calculated following
\begin{equation} \label{E:phistat}
    \Delta_\text{Static}(\rvec) = \Delta_\text{NCPA}(\rvec) + \sum_{k=1}^{n_\text{m}} \boldsymbol{\mu}_\text{Stat}(k)\boldsymbol{M}_k(\rvec),
\end{equation}
where $\boldsymbol{\mu}_\text{Stat}(k)$ and $\boldsymbol{M}_k$ are respectively the coefficient and 2D shape of the k$^\text{th}$ mode over $n_\text{m}$ modes. To include segment piston errors, $\boldsymbol{M}_k$ must be defined as the pattern formed by the k$^\text{th}$ segment (36 in total for the Keck pupil) and $a_k$ is the piston value in meters. Eventually, one may estimate both the atmospheric parameters and static coefficients and we illustrate such a joint estimation on Keck data in Sect. \ref{SS:piston}.

Finally, the description of the AO residual spatial filter $\tilde{k}_\text{AO}$ in Eq. \ref{E:otfdot} relies on two assumptions, which are, (i) the exposure is infinitely long, and (ii) the residual phase is a Gaussian statistical process. Thus, $\tilde{k}_\text{AO}$ is fully described by the residual phase power spectrum density (PSD) as follows \citep{Roddier1981}:
\begin{equation}
\begin{aligned}
    &\tilde{k}_\text{AO}(\rhol) = \exp\para{-0.5\times \dphi{\text{AO}}(\rhovec,\lambda)},\\
    & \dphi{\text{AO}}(\rhovec,\lambda) = 2\times\para{\cov{\text{AO}}(0,\lambda) - \cov{\text{AO}}(\rhovec,\lambda)},\\
    & \cov{\text{AO}}(\rhovec,\lambda) = \mathcal{F}\cro{\psd{\text{AO}}(\boldsymbol{k})},
\end{aligned}
\end{equation}
where $\dphi{\text{AO}}$, $\cov{\text{AO}}$, and $\psd{\text{AO}}$ are the residual phase structure function, \ccc{autocovariance function,} and PSD respectively, $\boldsymbol{k}$ is the spatial frequencies vector and $\mathcal{F}[x]$ is the 2D Fourier transform of $x$. In \citet{Fetick2019_Moffat_aa}, the PSD is described as a split function to separate the AO-corrected spatial frequencies from the uncorrected frequencies that follow a Kolmogorov's $-11/3$ power law \citep{Kolmogorov1941b}
\begin{equation}
    \psd{\text{AO}}(\boldsymbol{k}) = \left\lbrace 
    \begin{aligned}
    & \mathcal{M}(\boldsymbol{k}, A, \alpha_x, \alpha_y, \theta, \beta) +C \quad \text{for } k \leq k_\text{AO}\\
    & 0.023 \rz^{-5/3} k^{-11/3} \quad \text{for } k > k_\text{AO}
    \end{aligned}
    \right.,
\end{equation}
where $k = \abs{\boldsymbol{k}}$, $\rz$ is the Fried parameter \citep{Fried1966}, $k_\text{AO}$ the equivalent AO cut-off frequency beyond which the AO correction no longer occurd, $C$ a constant value and $\mathcal{M}$ an asymmetric Moffat function \citep{Moffat1969} that depends on five shape parameters $A, \alpha_x, \alpha_y, \theta$, and  $\beta$ and on the spatial frequency vector $\boldsymbol{k} = (k_x,k_y)$ as follows,
\begin{equation}
\label{E:psd}
\begin{aligned}
&\mathcal{M}(\boldsymbol{k}, A, \alpha_x, \alpha_y, \theta, \beta) =\\ &\dfrac{\psi\times A}{(1 + (k_x\cos(\theta) + k_y\sin(\theta))^2/\alpha_x^2 + (k_y\cos(\theta) - k_x\sin(\theta))^2/\alpha^2_y)^\beta},
\end{aligned}
\end{equation}
where $\psi$ is a normalization factor that ensures that the integral of the Moffat function over the AO-corrected area is given by $A$: \citep{Fetick2019_Moffat_aa}
\begin{equation}
    \psi = \dfrac{\beta-1}{\pi\alpha_x\alpha_y}\dfrac{1}{(1 - (1+ k_\text{AO}^2/(\alpha_x\alpha_y))^{1-\beta}}.
\end{equation}
Contrary to the classical use of a Moffat function in astronomy \citep{Trujillo2001,Moffat1969}, we stress that the Moffat function used in the present model serves in the description of the PSD and not the focal-plane PSF. This latter is deduced by the inverse Fourier transform of the OTF given in Eq. \ref{E:otfdot}. Consequently, the PSF FWHM is actually driven by parameters $A$ and $\rz$ rather than $\alpha_x, \alpha_y$, and $\beta$. In order to really capture the physical meaning of this model, below we present a description of each of the seven parameters that the model relies on and their impact on the PSF:
\begin{itemize}
    \item $\rz$ constrains the uncorrected PSD that corresponds to the PSF wings, that is, the part of the PSF that remains untouched by any AO correction. For a system with $n_\text{act}\times n_\text{act}$ actuators to compensate for the incoming wavefront, this breaking occurs in the focal plane at approximately $n_\text{act}\times\lambda/(2D)$, with $D$ the primary mirror diameter and for a Nyquist-sampled PSF \citep{Roddier1999}.
    \item $C$ is a constant value that \ccc{changes the gap between the AO-corrected part and the uncorrected high-spatial frequencies. }
    This parameter plays a major role near the AO cutoff in both the PSD and PSF planes, and compares to the aliasing error that contaminates wavefront measurements due to the WFS discrete sampling \citep{Bond2018,Correia2014,Jolissaint2010,Rigaut1998}.
    \item $A$ is the total energy in nm$^2$ contained in the Moffat PSD model (constant $C$ not included) and thus characterizes the AO residual wavefront error that connects to the PSF SR, that is, the intensity of the PSF peak compared to the diffraction-limit scenario.
    \item $\alpha_x,\alpha_y$, and $\theta$ govern the elongation and skewness of the PSD as well as the direction of the elongation thanks to the angle $\theta$. The PSD FWHM is proportional to $\alpha_x,\alpha_y$ parameters; for the same amount of energy, a PSD with a larger FWHM indicates that the PSD flattens and the correction homogenizes across spatial frequencies.
    \item $\beta$ represents the asymptotic slopes of the PSD at large spatial frequencies within the AO correction radius. From Fourier analysis of AO performance \citep{Bond2018,Correia2014,Jolissaint2010,Rigaut1998}, we know  that the PSD pattern introduced by the wavefront measurement noise follows a $k^{-2}$ power law \ccc{for WFSs sensitive to the first order derivative of the wavefront such as the Shack-Hartmann }. Asymptotically, we would have $\beta=1$ for a Shack-Hartmann based AO system if the measurement noise is the only error in the wavefront reconstruction process. Moreover, in the situation of an AO system correcting all atmospheric aberrations with the same relative level (which does not occur as the sensitivity of the wavefront sensor is aberration-dependent \citep{Fauvarque2016}), we would obtain $\beta = 11/6\simeq 1.83$ according to the von K\'arm\'ann expression of the atmospheric PSD \citep{Karman1948}. Nevertheless, larger values of $\beta$ can be observed. Indeed, for very large values of $\beta$, the Moffat distribution converges towards a Gaussian shape \citep{Trujillo2001}, as does the PSF consequently. We expect this behavior with poor tip-tilt correction for instance or in the presence of strong telescope wind-shake or telescope/dome vibrations. As a summary, $\beta$ should usually range between 1 and 1.83 for nominal AO correction but can reach higher values in the presence of non-atmospheric aberrations.
\end{itemize}

\begin{figure*}[h!]
    \centering
    \includegraphics[width=0.47\linewidth]{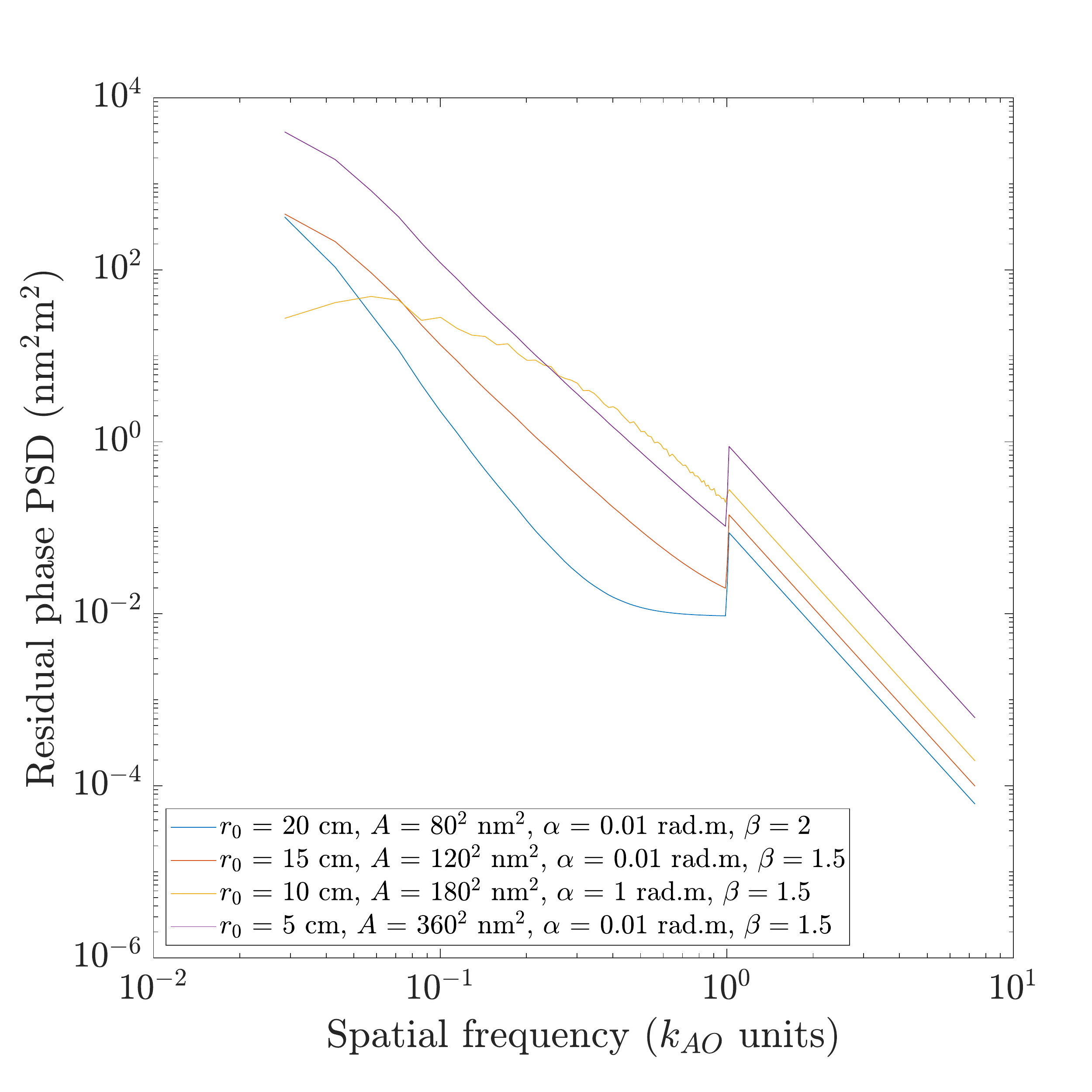}
    \hspace{0.02\linewidth}
     \includegraphics[width=0.47\linewidth]{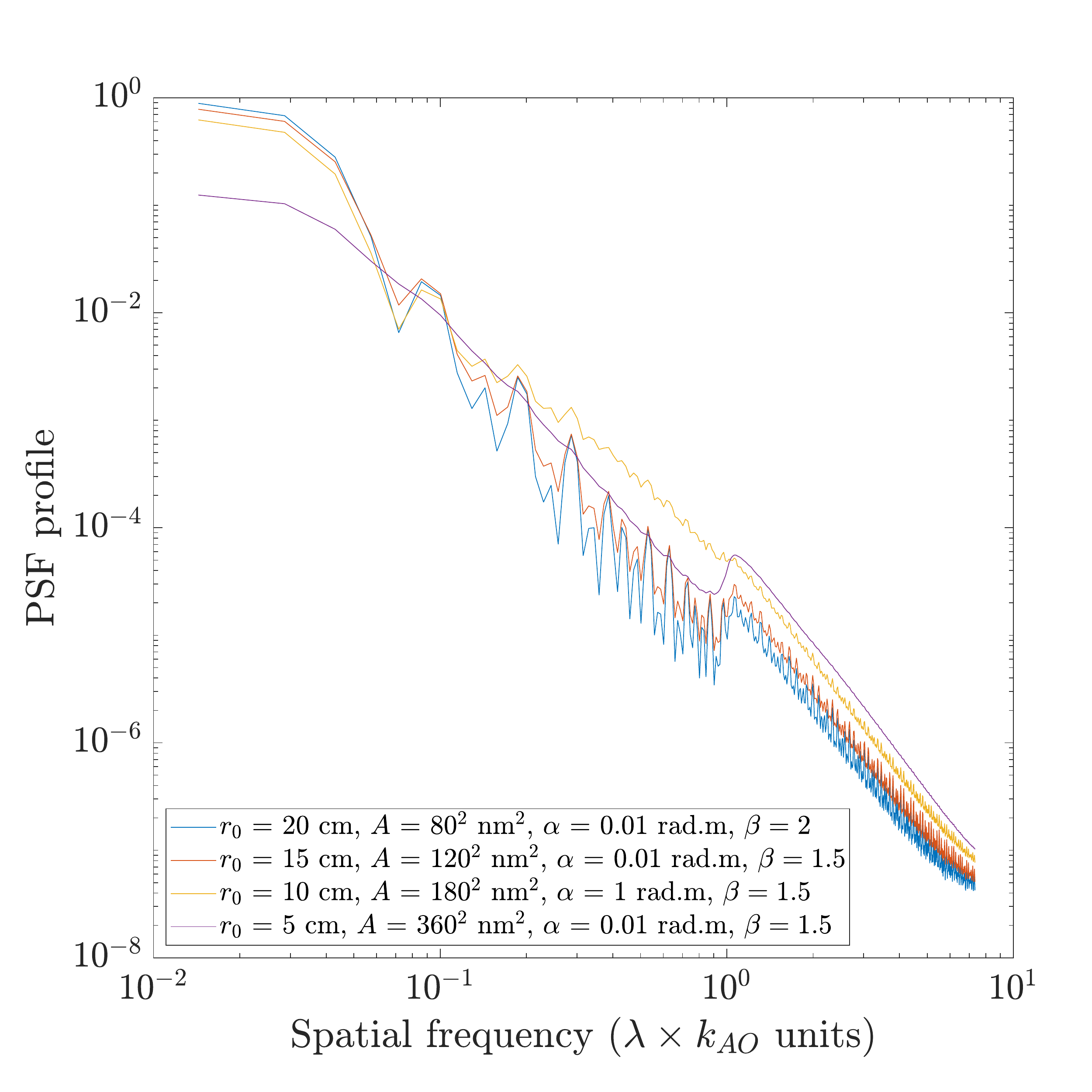}
    \caption{Azimuthal average in log-log scale of the PSD (left) and the PSF (right) obtained with four different settings for the model parameters, assuming that the PSD is symmetric, i.e. $\alpha_x=\alpha_y=\alpha$ and $\theta=0$. No static aberrations were added here and the parameter $C$ was set to $C = 10^{-4}$ rad$^2$m$^2$. The PSF maximum is normalized to the SR value.}
    \label{fig:psds}
\end{figure*}

We illustrate in Fig. \ref{fig:psds} azimuthal profiles of PSDs and corresponding PSFs for various sets of model parameters. This figure shows clearly how $\rz$ drives the PSF halo and how the parameter $A$ increases the area underneath the PSD curve, which corresponds to the residual wavefront error. We also observe that a larger value of $\beta$ sharpens the PSD. Knowing that the total energy remains constant, this situation corresponds to less efficient low-order modes compensation. Finally, a larger value of $\alpha$ increases the PSD FWHM and distributes more energy over the highest AO-correct modes. 
The shape of the PSF is mainly conditioned by parameters $A$ and $\rz$, which are the two most significant parameters of this model. Parameters $C$, $\alpha$, $\theta$, and $\beta$ are secondary parameters that shape the PSD in order to more accurately reproduce the PSF structure in various observing conditions and especially in the presence of sub-optimal AO control or non-atmospheric aberrations. Henceforth, this model is a combination of a PSF determination tool and an AO diagnosis facility gathered up together into a single and parsimonious analytical framework.

\section{Model versus reality: statistical validation}
\label{S:stat}

\subsection{Overview of data}
\label{SS:data}

We aim to provide the most exhaustive on-sky validation of the PSF analytical and parsimonious model proposed by \citet{Fetick2019_Moffat_aa}. To achieve this goal, we collected 4812 PSFs obtained from different observatories, AO systems, AO modes and spectral bands covering optical and NIR wavelengths. \ccc{The GALACSI/MUSE and GeMS/GSAOI instruments deliver simultaneous PSFs at different wavelengths and field positions respectively. If we consider a single PSF for each observation they produce, we obtain a total of 1880 PSFs. However, although the realizations of atmospheric residual wavefront are not independent from a PSF to another within the same observation, these data allow to test the model in the presence of chromatic and field-dependent instrumental aberrations. In order to conserve the same fitting process for all data, we have fitted each PSF with the same starting point and regardless the results obtained on another PSF acquired during the same observation.} A description of each data set is given below
\begin{itemize}
\item[$\bullet$] \textbf{SPHERE/ZIMPOL@VLT}.SPHERE is a facility at the VLT that was installed in 2015 and relies on an extreme AO (EXAO) system (Sphere Ao for eXoplanets Observations; \citealt{Fusco2016_SAXO,Sauvage2015}) to deliver a very efficient atmospheric correction level to three science instruments, which are ZIMPOL \citep{Schmid2018}, IRDIS \citep{Dohlen2008}, and IFS \citep{Claudi2014}. This set of data is composed of two subsets of PSFs from SPHERE/ZIMPOL. The first (26 PSFs) was obtained from observations of NGC 6121 in 2019 \citep{Massari2020,BeltramoMartin2020_ZIMPOL_PSFR} with the ZIMPOL V-filter (central wavelength 554\,nm, width 80.6\,nm) and a pixel scale of 7.2\,mas/pixel in the context of technical calibrations \footnote{ESO program ID of observations: 60.A-9801(S)} granted after the 2017 ESO calibration workshop (see \url{http://www.eso.org/sci/meetings/2017/calibration2017}). The second set of 18 PSFs was acquired in 2018 with the N\_R-filter (central wavelength 645.9\,nm, width 56.7\,nm) and a pixel scale of 3.6\,mas/pixel  during the ESO Large Program (ID 199.C-0074, PI P. Vernazza) \citep{Vernazza2018}. The data were reduced using the SPHERE Data and Reduction Handling pipeline (DRH) to extract the intensity image, subtract a  bias frame, and correct for the flat-field. These data are particularly useful for testing the model in optical wavelengths and under strong atmospheric residual regime.
\item[$\bullet$] \textbf{SPHERE/IRDIS@VLT}. 
A total of 237 PSFs were obtained from the SPHERE Data Centre client \citep{Delorme2017} using the Keyword \emph{Frame type} set to IRD\_SCIENCE\_PSF\_MASTER\_CUBE. These PSFs were acquired over the last five years with the N\_ALC\_YJH\_S APLC \citep{Vigan2010MNRAS} during PSF calibration and with a pixel scale of 12.5 mas/pixel. These data were collected using the dual band filters DB\_H23, and DB\_K12, and they are useful for testing the model under very high SR regime and for validating the LWE retrieval presented in Sect. \ref{SS:lwe}.
\item[$\bullet$] \textbf{Keck AO/NIRC2@Keck II}. We obtained 355 PSFs using the narrow field mode of NIRC2 with a pixel scale of 9.94 mas/pixel and using the Fe II and K cont filters. The Keck AO system on the Keck II telescope was operated in single conjugated AO (SCAO) mode using a natural \citep{Wizinowich2000} or an on-axis laser \citep{Wizinowich2006} guide star.  These PSFs were obtained during PSF reconstruction engineering nights in 2013 and 2017 \citep{Ragland2018_PSFR,Ragland2016} and such data are especially useful to validate the model under the influence of remaining piston cophasing errors, as we discuss in Sect. \ref{SS:piston}.
\item[$\bullet$] \textbf{SOUL/LUCI@LBT} 
These data were delivered in 2020 from the to two LUCI NIR spectro-imagers assisted with the pyramid SCAO (PSCAO) SOUL AO system \citep{Pinna2016} driven by a pyramid WFS \citep{Ragazzoni1999b}. These PSFs were acquired using the H filter (1.653\,$\mu$m) and with a sampling of 15\,mas/pixel. Although few (11 for our purpose) data sets have been obtained so far in comparison to others systems, these PSFs pave the way to a demonstration of PSF determination strategies in the presence of a pyramid WFS, which will be the baseline for the SCAO mode of HARMONI \citep{Thatte2017}, MICADO \citep{Davies2016} \ccc{and METIS \citep{Hippler2019}} on the future 39\,m Extremely Large Telescope (ELT). 
\item[$\bullet$] \textbf{CANARY/CAMICAZ@WHT} CANARY \citep{Gendron2010,Myers2008} was designed as a pathfinder for demonstrating the reliability and robustness of the multi-object AO (MOAO) concept proposed for assisting very large field (>1') multi-object spectrographs, such as MOSAIC for the ELT \citep{Hammer2016}. Thanks to 26 nights of commissioning, tests, and validation, we collected 1268 PSFs in H band and with a sampling of 30\,mas/pixel using the NIR detector CAMICAZ \citep{Gratadour2014}. These data were obtained  during phase B \citep{Martin2017,Morris2014,Morris2013,Martin2013}, which was dedicated to the demonstration of MOAO relying jointly on four (Rayleigh) laser guide star (LGS) and up to three natural guide star (NGS) in a 1' FOV to perform the tomography using the Learn \& Apply technique \citep{Laidlaw2019,Martin2016L3S,Vidal2010}. Among this quite large set of data, we have 522 PSFs in SCAO mode, 128 PSFs in ground layer AO (GLAO) mode and 618 MOAO PSFs. Such an archive is useful for testing the model on a 4.2\,m class telescope and on tomographic and laser-assisted AO-corrected-PSFs.
\item[$\bullet$] \textbf{GALACSI/MUSE NFM@VLT} MUSE is the ESO VLT second-generation wide-field integral field spectrograph operating in the visible \citep{Bacon2010},  covering a simultaneous spectral range of 465-930\,nm and assisted by the ESO Adaptive Optics Facility (AOF, \citealt{Oberti2018,Arsenault2008}) including the GALACSI module \citep{Strobele2012}. We focused our analysis on the narrow field mode (NFM) of MUSE, which delivers a laser tomography AO (LTAO)-corrected field covering a 7.5" $\times$ 7.5" FOV, providing near-diffraction-limited images with a sampling of 25\,mas/pixel. These PSFs were obtained during the commissioning phase in 2018 and are particularly useful for analyzing the model outputs with respect to the wavelength and demonstrating that the model also complies with under-sampled PSFs. Also, we performed a spectral binning to reach a spectral width of 5\,nm (91 PSFs per cube if we remove the notch filter wavelengths), resulting in a total of 1986 PSFs.
\item[$\bullet$] \textbf{GEMS/GSAOI@GEMINI} The Gemini South Adaptive Optics Imager (GSAOI) is a NIRd camera that benefits the correction provided by the Gemini Multi-conjugate Adaptive Optics (MCAO) System (GeMS, \citealt{Neichel2014_GEMINI,Rigaut2014}) on Gemini South. It delivers near-diffraction-limited images in the 0.9 - 2.4\,$\mu$m wavelength range in a large FOV of 85"$\times$85". From the Gemini archive \citep{Hirst2017}, we  assembled a catalog of 911 isolated and nonsaturated PSFs extracted from 60 images of Trumpler 14 acquired in 2019 and using the J, H, K, and Br$\gamma$ filters with a pixel scale of 20\,mas/pixel (P.I. M. Andersen, observation program: GS-2019A-DD107). These data are particularly useful for investigating model parameters variations across a MCAO-corrected FOV.
\end{itemize}

In total, we created a dictionary of 4812 PSFs covering several observatories, AO correction types, optical and NIR wavelengths, as summarized in Table \ref{tab:data}. Having this diversity of data is important for spanning the full range of possible AO correction levels and assessing which conditions must be met to achieve an accurate PSF representation.
\begin{table*}[h!]
    \centering
    \begin{tabular}{|c|c|c|c|c|}
    \hline
    SYSTEM & AO MODE & $\lambda$ range ($\mu$m) & \# PSFs & SPECIFICITY\\
    \hline
     SPHERE/ZIMPOL & EXAO & 0.55 to 0.65 & 44 & NGS, High Order AO system   \\
    \hline
    SPHERE/IRDIS & EXAO & 1.65 to 2.2 & 237 & NGS-AO residual NCPA/LWE \\
    \hline
     KECK AO/NIRC2 & SCAO & 1.65 to 2.2 & 355 & Natural and Laser assisted AO / Segmentation  \\
    \hline
     SOUL/LUCI & PSCAO & 1.65 & 11 & High Order Pyramid WFS  \\
    \hline
     CANARY/CAMICAZ & MOAO & 1.65 & 1268 & 4.2m pupil \\
    \hline
    GALACSI/MUSE & LTAO &  0.46 to 0.93 & 1986 &  Spectro-imager and undersampled image \\
    \hline
    GEMS/GSAOI & MCAO & 1.12 to 2.2 & 911 & Large fov \\
    \hline
    \end{tabular}
    \caption{Summary of PSFs obtained and processed for the analysis presented in this paper. In total, the model has been tested over 4812 PSFs. }
    \label{tab:data}
\end{table*}

\subsection{PSF fitting}

In order to fit the model over the image and retrieve the associated parameters, we followed the same strict process for each of the 4812 PSFs in our dictionary, which is described as follows
\begin{itemize}
    \item Define a model of the image including the PSD degrees of freedom (no static aberrations retrieval)  $\boldsymbol{\mu}_\text{AO}$ as well as four additional scalar factors $\delta_x$, $\delta_y$, $\gamma$, and $\nu$ in order to finely adjust the PSF position, flux, and constant background level,
    \begin{equation}
    \begin{aligned}
        &\hat{d}(\boldsymbol{\mu}_\text{AO},\gamma, \delta_x,\delta_y,\nu) =\\ &\gamma.\mathcal{F}\cro{\otf{\text{Static}}.\tilde{k}_\text{AO}(\boldsymbol{\mu}_\text{AO}).\exp\para{2i\pi\times(\rho_x\delta_x/\lambda+\rho_y\delta_y/\lambda)}} + \nu.
        \end{aligned}
    \end{equation}
    The image model is calculated over a given number of pixels that is 10\% larger than the on-sky images so as to mitigate aliasing effects due to the Fast Fourier Transform algorithm. The size of the on-sky image support is instrument-dependent and truncated in order to mitigate the noise contamination. When possible, we crop the on-sky image to conserve a FOV up to twice the AO cutoff.
    \item Define a criterion to minimize 
    \begin{equation}
    \begin{aligned}
        &\varepsilon(\boldsymbol{\mu}_\text{AO},\gamma, \delta_x,\delta_y,\nu) =\\ &\sum_{i,j} W_{ij}\cro{\hat{d}_{ij}(\boldsymbol{\mu}_\text{AO},\gamma, \delta_x,\delta_y,\nu) - d_{ij}}^2,
        \end{aligned}
        \label{E:crit}
    \end{equation}
    where $d_{ij}$ is the $(i,j)$ pixel of the 2D image and $W_{ij}$ the weight matrix defined by
    \begin{equation}
        W_{ij} = \dfrac{1}{\text{max}\left\lbrace d_{ij},0\right\rbrace +  \sigma^2_\text{ron}}.
    \end{equation}
    The weight matrix accounts for the noise variance, that is, both photon noise and read-out noise, and allows us to maximize the robustness of the fitting process \citep{Mugnier2004}.
    \item Perform the minimization using a nonlinear and iterative recipe based on the trust-reflective-region algorithm \citep{conn2000trust}.  We did not use specific regularization techniques on top of the weighting matrix as we have taken care of selecting good S/N images. 
\end{itemize}

\subsection{PSF model accuracy}

\begin{figure*}[h!]
    \centering
    \includegraphics[width=\linewidth]{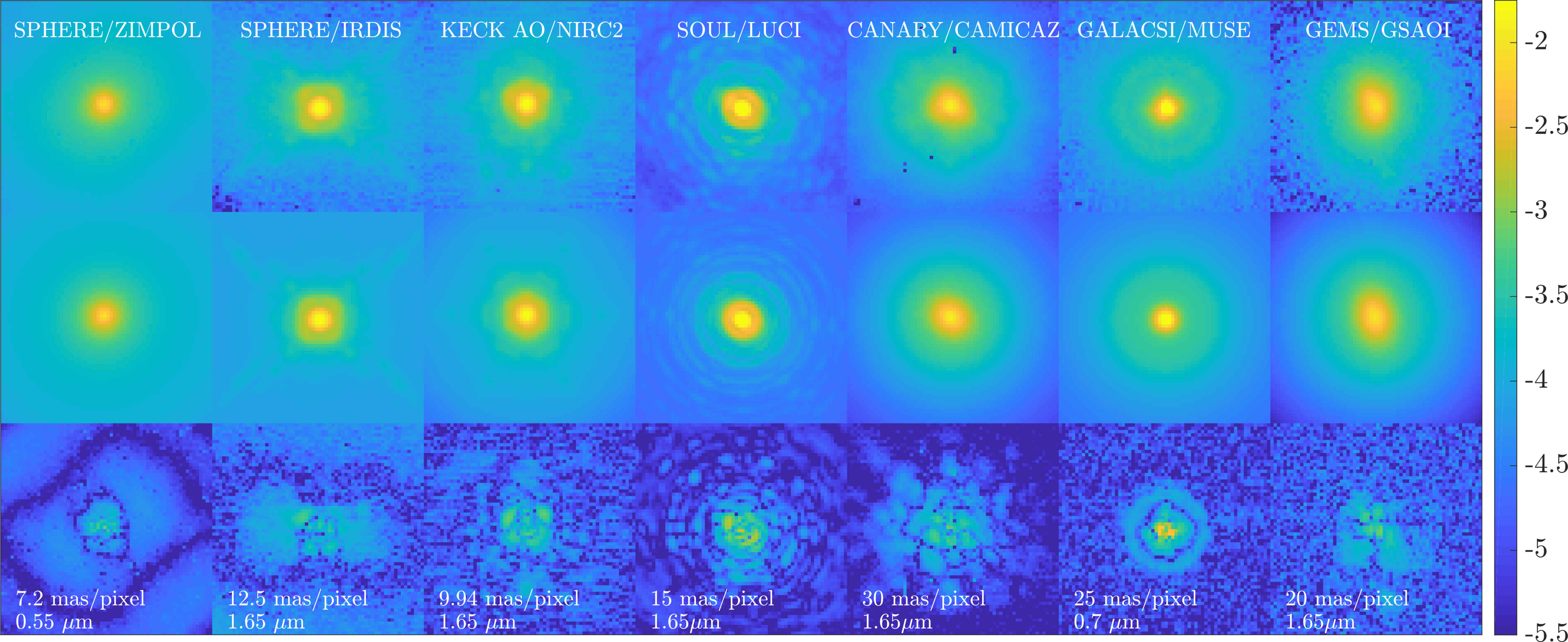}
    \caption{Two dimensional comparison in log scale and over 64$\times$ 64 pixels of (top) observed PSFs, (middle) fitted model, and (bottom) residual map for multiple types of AO systems and instruments working in either visible or NIR wavelengths. From left to right: SPHERE/ZIMPOL, SPHERE/IRDIS with the apodizer and the Lyot stop (no focal plane mask), Keck AO/NIRC2, SOUL/LUCI, CANARY/CAMICAZ in MOAO mode, GALACSI/MUSE NFM, and GEMS/GSAOI. All PSFs have been normalized to the sum of pixels. }
    \label{fig:overall_psfs_2D}
\end{figure*}

In Fig. \ref{fig:overall_psfs_2D}, we present a visual comparison of on-sky PSFs, best-match model and residual map for the seven instruments considered in this analysis. The model adapts to any kind of AO correction; the maps of residuals are visually similar to first order among all systems. On the Keck AO/NIRC2 image, we see some structures that are static speckles, probably introduced by a remaining cophasing error, and residual NCPA as illustrated in Sect. \ref{SS:piston}. On SOUL/LUCI and CANARY/CAMICAZ images, we also see a persistent pattern that can be explained by the presence of static aberrations not included in the model and the exposure time that was not sufficiently long (a few seconds of exposure) to average the atmospheric speckles.

\begin{figure*}
    \centering
    \includegraphics[width = 0.47\linewidth]{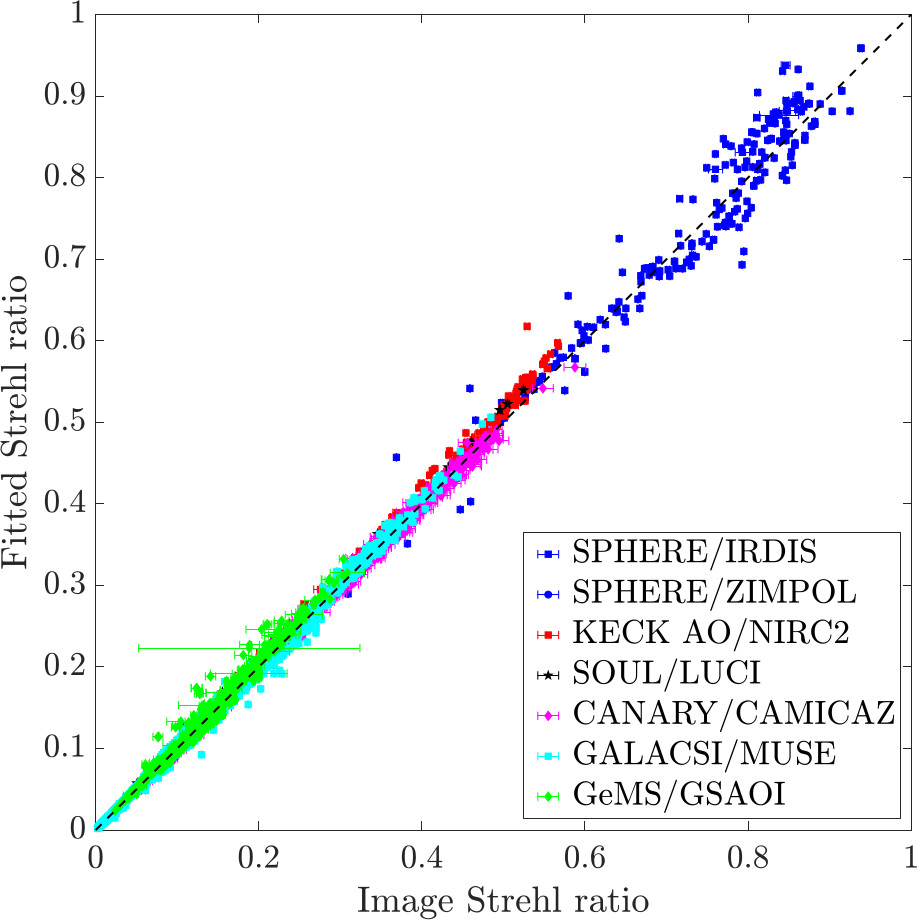}
    \hspace{0.02\linewidth}
    \includegraphics[width = 0.48\linewidth]{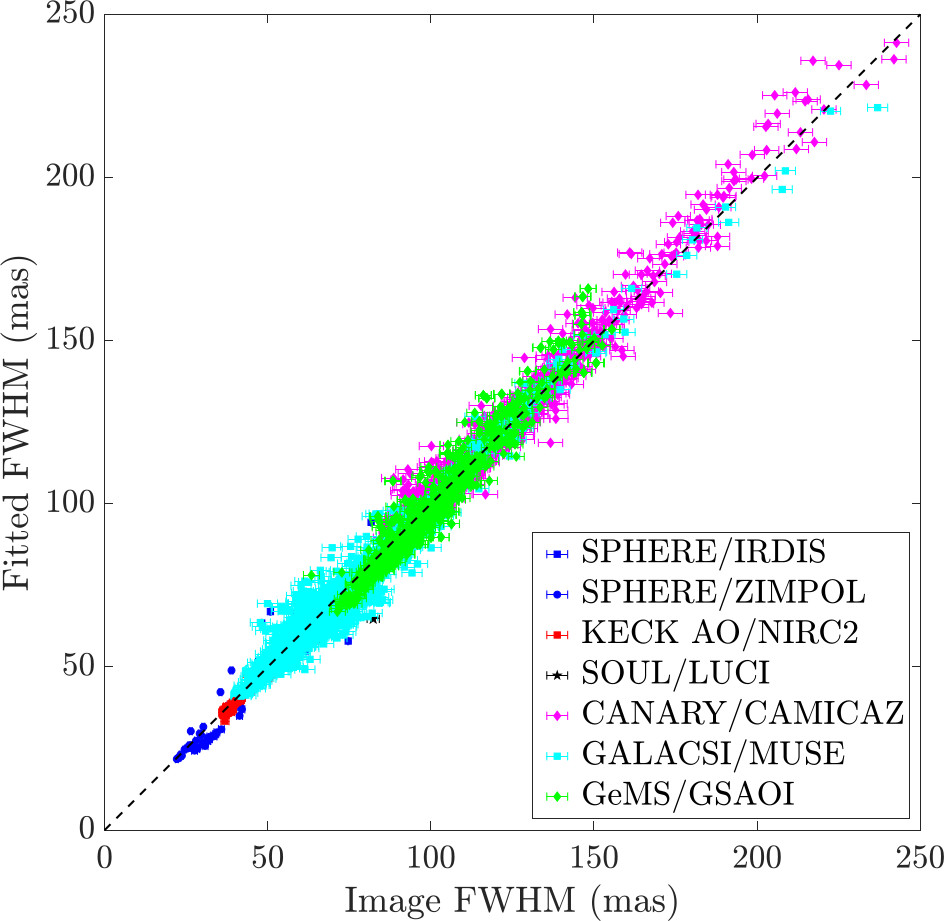}
    \caption{Image SR/FWHM versus the same metrics retrieved on the fitted image using the same estimation process and for the 4812 PSFs treated for this analysis. Error bars on the SR are obtained from calculations presented in \citep{Martin2016JATIS}. Error bars on the FWHM are given from the contour estimation on interpolated images that are oversampled by up to a factor four to quantify the FWHM more accurately.   }
    \label{fig:SRplot}
\end{figure*}

In Fig. \ref{fig:SRplot}, we illustrate the SR and FWHM obtained from the fitted model versus sky image-based estimates. The same algorithms (OTF integral for the SR, interpolation+contour for the FWHM) were used to calculate these metrics regardless of the nature of the data, either sky image or model. For all systems, we observe a remarkable correlation and a similar dispersion on both SR and FWHM and for all observing conditions, AO modes and AO correction levels. For very high SR values obtained with SPHERE/IRDIS, we start observing an overestimation of the SR suggesting that the model cannot fully retrieve some patterns in the image. At this level of correction, the instrumental contribution (residual NCPA, LWE for instance) of the PSF may dominate the PSF morphology while they are not included in the model, as for all 4812 handled data sets. Consequently, in order to guarantee an SR and FWHM accuracy at a few percent, it is not necessary to include a precise model of these instrumental defects for AO systems delivering SR up to $\sim$80\%. We confirm this assumption in Sect. \ref{S:static}. By calculating the relative difference ($(x_\text{model}-x_\text{sky})/x_\text{sky}$) on SR and FWHM values over all the 4812 data sets, we measure a bias and a standard deviation (std) value of 0.7\% and 4.0\% on the SR estimation and -0.8\% and 4.6\% on the FWHM estimation.  These numbers indicate that there is a marginal performance overestimation of 1\% from the model (larger SR, lower FWHM), which  fits the measurements uncertainty envelopes. As a result, this model achieve a PSF recovery at a 4\% level.

Table \ref{tab:statSR} provides statistical results of the estimated SR and FWHM PSFs for all instruments, including the median values  for all observations and the Pearson correlation factor. As suggested by Fig. \ref{fig:SRplot}, there is no specific bias, except for SPHERE/IRDIS for reasons mentioned above, as well as for SOUL/LUCI owing to the small amount of data we have access to so far. Overall, we conclude that (i) the model becomes biased for very high SR observations (SR > 80\%), calling for the introduction of instrumental defects to improve the model accuracy, (ii) the SR is estimated with a 1-$\sigma$ precision of 1\,\%, and  (iii) the FWHM is estimated with a 1-$\sigma$ precision of 3\,mas, which correspond to approximately to one-fifth down to one-tenth the pixel scale depending on the instrument.

\begin{table*}[h!]
\ccc{
    \centering
    \begin{tabular}{|c|c|c|c|c|c|c|c|c|c|}
    \hline
    SYSTEM & $\lambda$ ($\mu$m) & \multicolumn{4}{c|}{SR (\%)} & \multicolumn{4}{c|}{FWHM (mas)}   \\
    \hline
    & & Median & bias & std & Pearson & Median & bias & std  & Pearson \\
    \hline
    \multirow{2}{*}{SPHERE/ZIMPOL} & 0.55 & 6.2 & 0.5 & 0.3 & 0.998 & 32& -4.2 & 0.9& 0.995 \\
    & 0.64 & 12.4 & 0 & 0.38 & 0.999 & 26 & 1 & 2.9& 0.96 \\
    \hline
    \multirow{2}{*}{SPHERE/IRDIS}& 1.67 & 61 & -0.2 & 2.0 & 0.98 & 52 & 0.2 & 1.4 & 0.90\\
    & 2.25 & 80.0 & 0.7 & 3.2 & 0.98 & 66 & 0.8 & 1.5 & 0.88\\
    \hline
    \multirow{3}{*}{GALACSI/MUSE} & 0.5 & 2.4 & -0.02 & 0.2 & 0.993 & 80 & -1.4 & 2.9 & 0.997 \\
    & 0.7 & 10.3 & 0.1 & 0.3 & 0.998 & 69 & -1.7 & 3.2 & 0.980 \\
    & 0.9 & 25.0 & 0.2 & 0.7 & 0.998 & 58 & 0.08 & 5.4 & 0.90 \\
    \hline
    \multirow{2}{*}{KECK AO/NIRC2} & 1.65 (NGS) & 39.4 & 1.1 & 0.9   & 0.998 & 37 & -0.9 & 0.8 & 0.994\\
    & 2.2 (LGS) & 22 & 0.7 & 0.6 & 0.999 & 70 & -1.9 & 1.3 & 0.995\\
    \hline
    SOUL/LUCI & 1.65 & 36.0 & 1.2  & 0.5 &0.990 & 55 & -2.5& 3.3 & 0.980\\
    \hline
    CANARY/CAMICAZ & 1.65 & 23.0 & 0.1 & 0.4& 0.999 & 115 & -0.1 & 3.7 & 0.993 \\
    \hline
    \multirow{3}{*}{GEMS/GSAOI} & 1.25 & 13.3 & 0 & 0.5 & 0.994 & 109 & 2.0 & 5.4 & 0.976\\
    & 1.64 & 7.6 & 0.4 & 0.5 & 0.990 & 98 & -4.0 & 3.2 & 0.98\\
    & 2.2 & 15.6 & 0.4 & 0.7 & 0.994 & 94 & -0.8 & 4.5 & 0.970\\
    \hline
    \end{tabular}
    \caption{Individual statistics per system and imaging wavelength on SR and FWHM estimates. The columns "Median" give median values estimated on images, while columns "bias", "std" and "Pearson" give the median of SR/FWHM error (in percent) and the Pearson correlation coefficient respectively.  }
    \label{tab:statSR}
    }
\end{table*}

\subsection{Exploitation of the model outcomes for diagnosing observing conditions and AO performance }

As discussed above, fitting the shape of the residual phase PSD allows to retrieve atmospheric parameters and AO performance. Therefore, the goals in this section are to (i) confirm that the output parameters $r_0$ and $A$ follow expected trends and give confidence in the retrieval process, and  (ii) provide statistics on $\alpha$, $\beta$ parameters to asses which values they should reach for a nominal AO correction and thus discriminate situations of sub-optimal AO correction. We have excluded the parameter $\theta$ from this analysis as it corresponds to a PSF orientation only.

\subsubsection{Primary parameters estimates}

The seeing is estimated from the PSF wings fitting relying on a Kolmogorov expression of the PSD. This measurement technique has proven to be robust \citep{Fetick2019_Moffat_aa} as it consists in determining a single parameter from a significant number of pixels. Thanks to the large redundancy across the pixels of the spatial signature that the algorithms is seeking out, this approach still gives meaningful results with moderate S/N in comparison to external profilers \citep{Fetick2019_Moffat_aa}. However, contrary to these latter, this focal-plane-based seeing determination includes all turbulence effects that contribute to impact the PSF wings, such as the dome seeing \citep{Lai2019,Conan2019_GMT}, which the external profilers are not sensitive to as they are apart from the dome. Consequently, estimating the $\rz$ from the focal-plane image allows to diagnose more accurately the AO performance in comparison to an external profiler.

The target here is to verify that the retrieved seeing values are consistent with what we know about the observing sites. The first verification we made is to analyze statistics on the seeing estimates presented in Table \ref{tab:seeing}. Given that the Keck data were acquired in February, August, and September seasons, the median seeing at Mauna Kea is consistent with the literature \citep{Ono2016,Miyashita2004}. Similar observations can be drawn for La Palma by comparing to either CANARY telemetry data \citep{Martin2016L3S} or the RoboDIMM \citep{OMahony2003}. We also find  consistency with analysis by \citet{Masciadri2014} for the Paranal site and by \citet{Tokovinin2006} for Cerro Pach\'on.

\begin{table}[h!]
    \centering
    \begin{tabular}{|c|c|c|}
    \hline
    Observing site     & median seeing & 1-$\sigma$ std \\
    \hline
    Mauna Kea& 0.60"& 0.13" \\
    \hline
    Cerro Pach\'on & 0.67" & 0.09" \\
    \hline
    Paranal & 0.85" & 0.15" \\
    \hline
    La Palma & 0.90" & 0.22" \\
    \hline
    \end{tabular}
    \caption{Median and 1-$\sigma$ standard-deviation of seeing values retrieved from the PSF-fitting process. Seeing values are given at zenith and 500\,nm.}
    \label{tab:seeing}
\end{table}

\begin{figure*}[h!]
    \centering
    \includegraphics[width=0.47\linewidth]{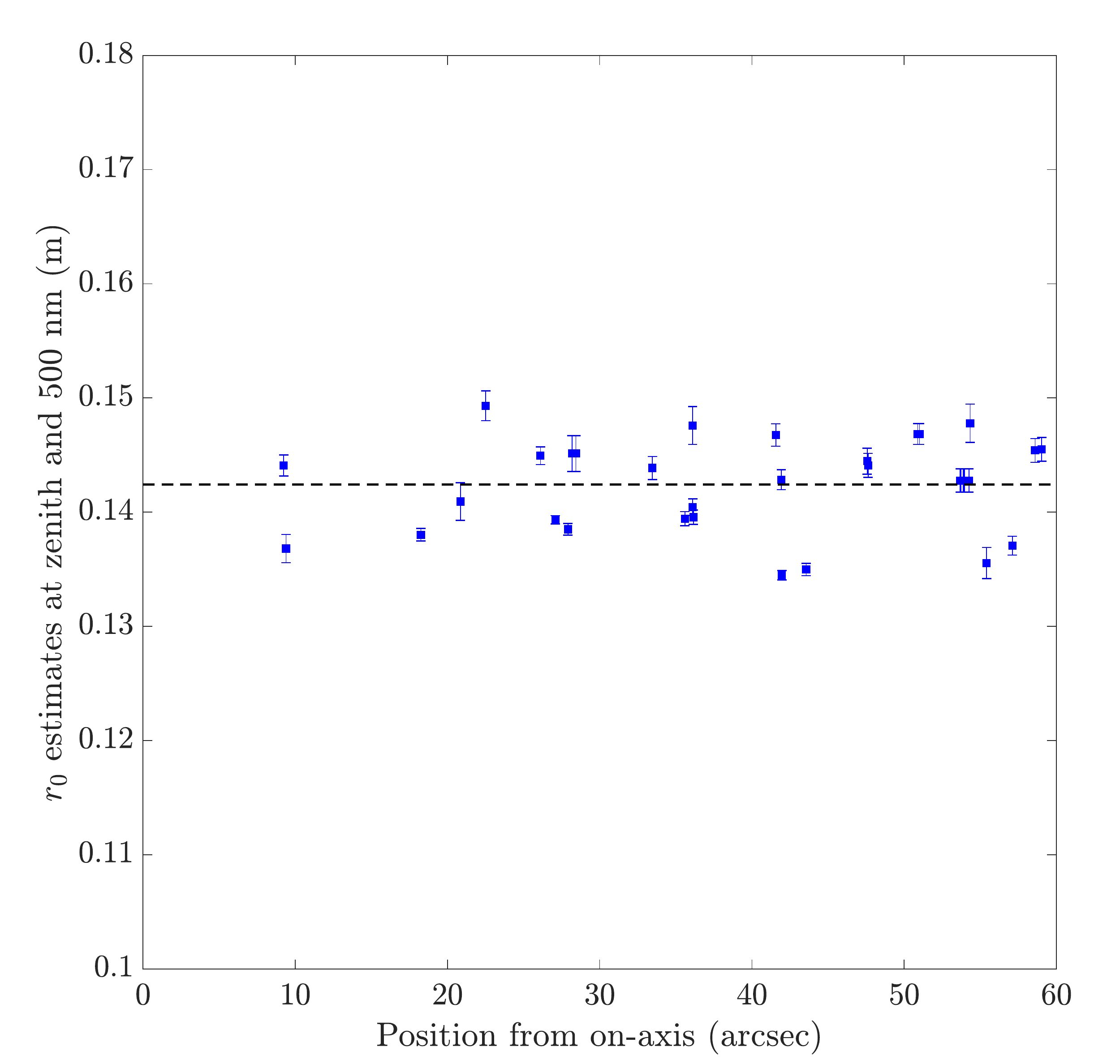}
    \hspace{0.02\linewidth}
    \includegraphics[width=0.47\linewidth]{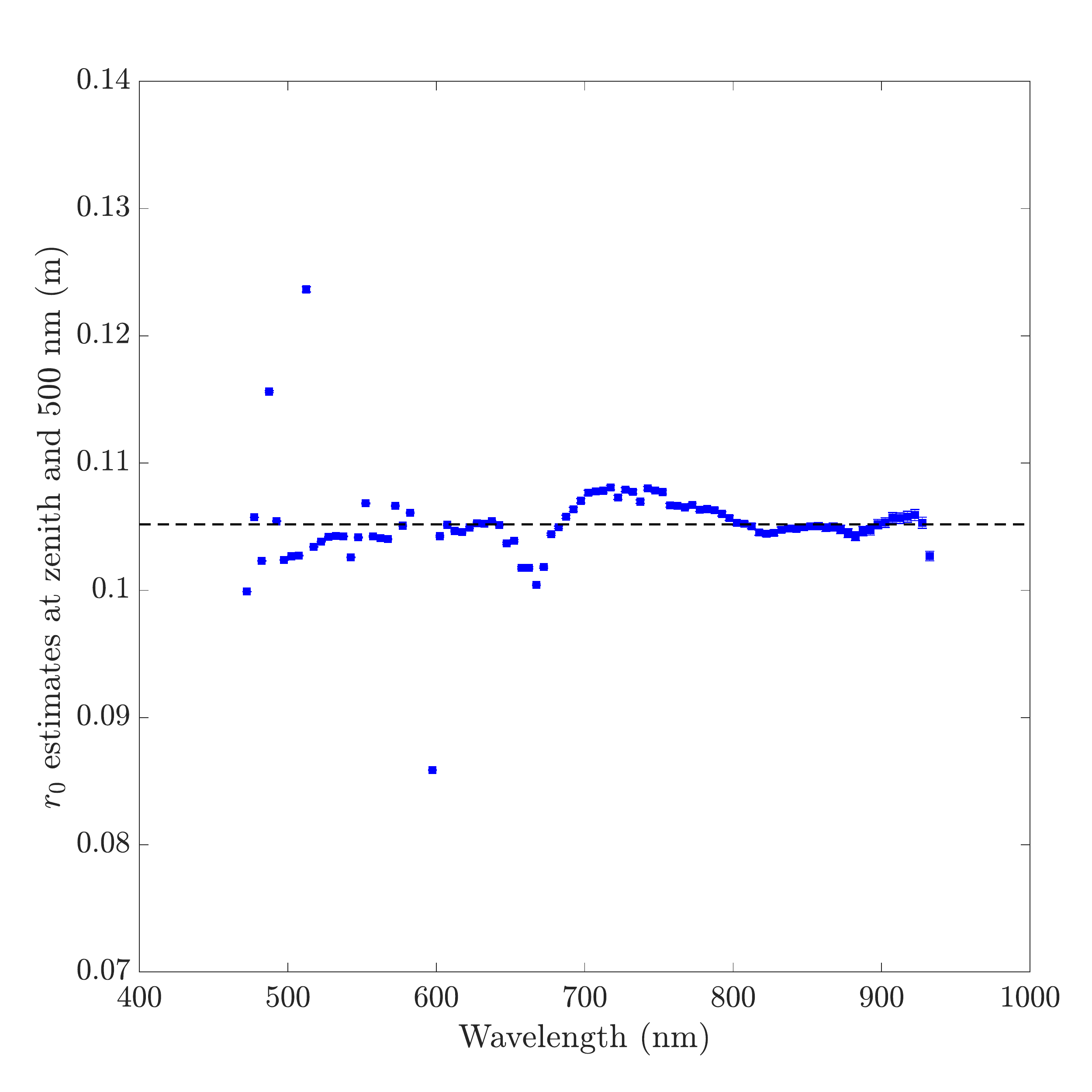}
    \caption{ $r_0$ estimates at zenith and 500\,nm with respect to the GEMS/GSAOI PSF field position (left) and with respect to the PSF GALACSI/MUSE wavelength (right).}
    \label{fig:r0}
\end{figure*}

In addition, we highlight in Fig. \ref{fig:r0} the $r_0$ estimates at zenith and 500\,nm obtained on a single image of GeMS/GSAOI and a single cube of GALACSI/MUSE. For the GEMS/GSAOI case, we have $r_0$ measurements obtained from several PSFs distributed over the field. Given that $r_0$ is assessed from the PSF wings that contain the non-AO-corrected spatial frequencies, we expect $r_0$ estimates to be independent of the PSF position in the FOV. Similarly,  we multiplied $r_0$ estimates obtained on GALACSI/MUSE images by a factor $(500/\lambda_c)^{6/5}$, where $\lambda_c$ is the central wavelength of the considered image in the data cube, in order to disable the theoretical dependency of $r_0$ with respect to $\lambda$.

Using our model, we retrieve $r_0$ as 14.2\,cm$\pm$0.4 and 10.4\,cm$\pm$0.2\,cm for GEMS/GSAOI and GALACI/MUSE, respectively. The achieved 3\% 1-$\sigma$ error of our estimates shows that this method is robust and precise compared to telemetry-based techniques that typically reach 10\% precision \citep{Jolissaint2018}.

\begin{figure*}[h!]
    \centering
    \includegraphics[width=0.47\linewidth]{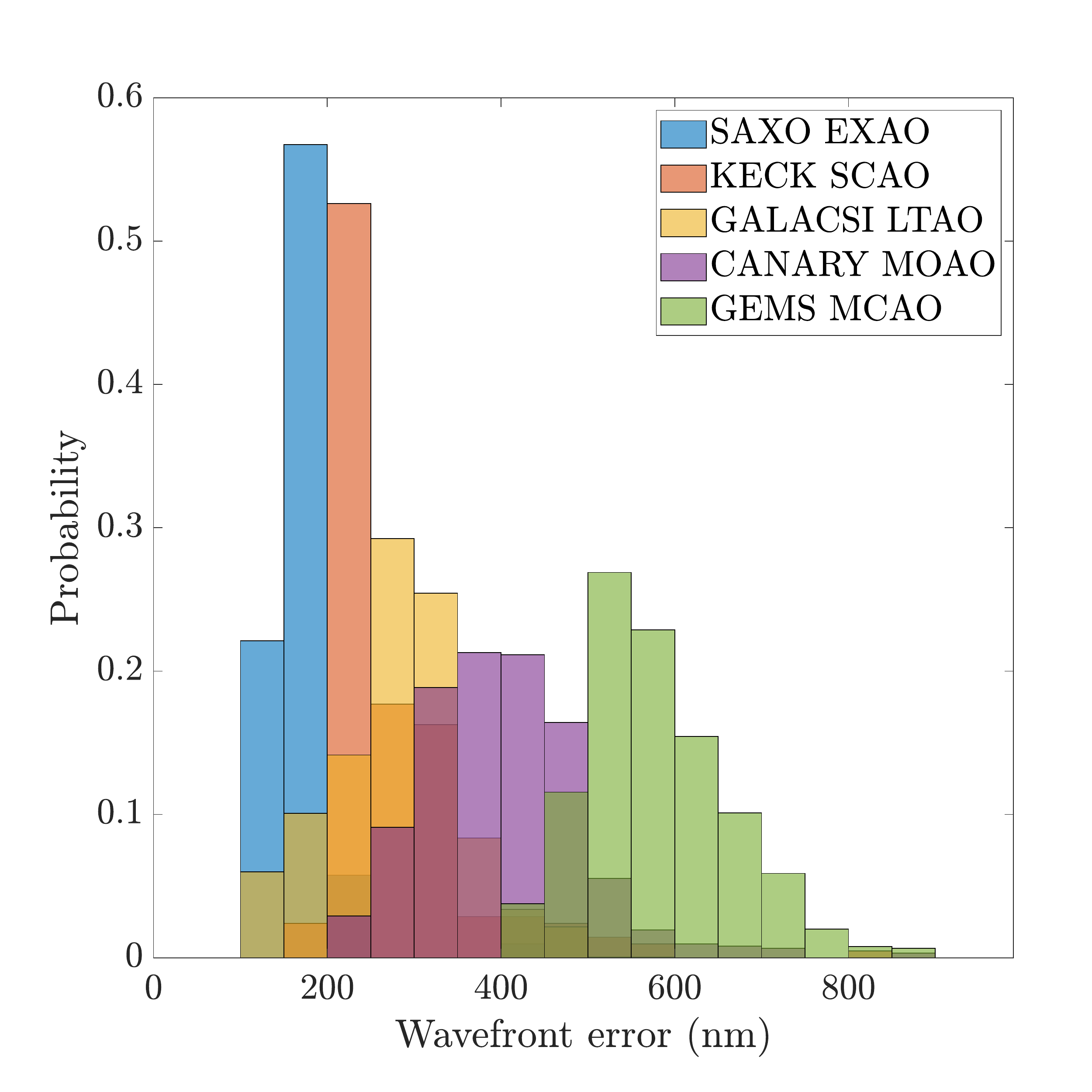}
    \hspace{0.02\linewidth}
     \includegraphics[width=0.47\linewidth]{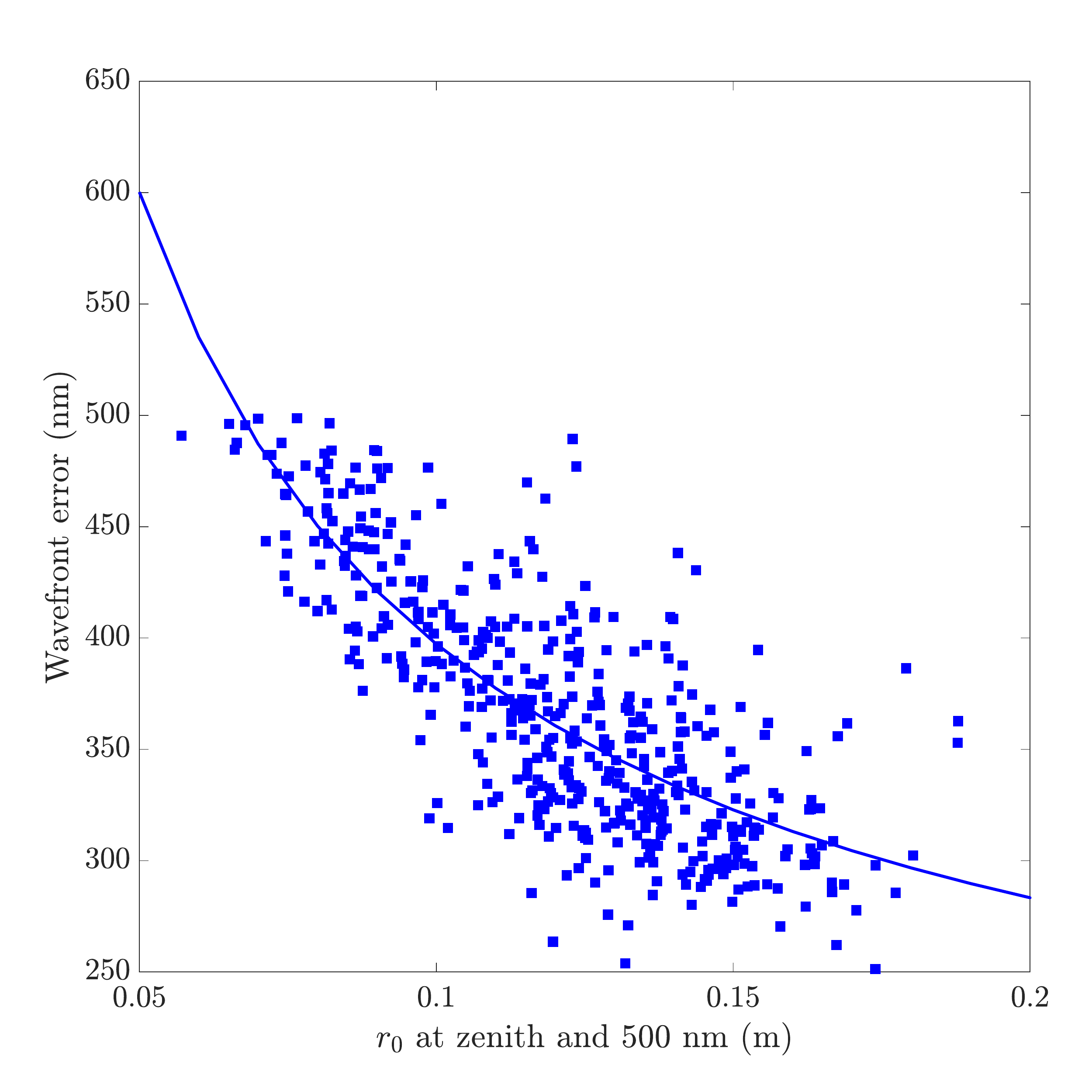}
    \caption{\textbf{Left:} Histograms for the $A$ parameter for each AO system. \textbf{Right:} Wavefront error obtained on CANARY SCAO mode with respect to the retrieved $r_0$. The solid line gives the trend in $r_0^{-5/6}$.}
    \label{fig:sig2}
\end{figure*}

 For completeness, Fig. \ref{fig:sig2} shows histograms for the $A$ parameter (the wavefront error) and the function $A = $f$(\rz)$ for the specific case of CANARY working in SCAO (i.e., the largest SCAO PSF dictionary we have). We clearly see that the wavefront standard-deviation error follows a $\rz^{-5/6}$ law as we expect from the von K\'arm\'an PSD \citep{Karman1948}, which is proportional to $\rz^{-5/3}$ for a SCAO system whose on-axis PSF is not influenced by the atmospheric turbulence profile. This illustrates the agreement between the different retrieved parameters.

Moreover, \ccc{the histogram of $A$ estimates given in Fig. \ref{fig:sig2}} shows that the retrieved wavefront errors vary within a meaningful range: the SPHERE AO system performs better than others, as expected from such an extreme AO system, with a median of 140\,nm, which includes all the aberrations that blur the AO-corrected part of the PSF. \ccc{The present SPHERE results gather PSFs obtained with SPHERE/ZIMPOL, acquired with a AO loop running at low frequency (300\,Hz, $m_V \simeq 11$), and SPHERE/IRDIS fir which 25\% were contaminated by a strong LWE. Therefore, this wavefront error of 140 nm seems consistent with the literature \citep{Sauvage2016}}. The Keck AO system reaches 230\,nm, which also complies with \citet{Wizinowich2012}. From GALACSI/MUSE NFM images, we retrieve 285\,nm of wavefront error, which agrees with the \citet{Oberti2018} analysis Moreover, the GALACSI system was not yet fully optimized to operate in LTAO and a recent acquisition in 2019 already showed improvements on SR. For CANARY/CAMICAZ,  we obtain a wavefront errors of 320\,nm, 396\,nm and 417\,nm in SCAO, MOAO, and GLAO mode respectively, which compares well with \citet{Martin2017} and \citet{Vidal2014}. Finally, GeMS/GSAOI data unveil a median wavefront error of 570\,nm from PSFs extracted from the field at 20" up to 60" off-axis, which corresponds to a SR of 7.8\% at the edges of the field and complies with analysis of the performance of GeMS \citep{Neichel2014_GEMINI,Rigaut2012}.

We stress that this value of wavefront error is not deduced from the image SR but directly from the integral of the PSD, which can be determined from the model parameters \citep{Fetick2019_Moffat_aa}. This corresponds to the real mathematical definition of the wavefront error and is not influenced by the Mar\'echal approximation \citep{Parenti1994} that is biased at low SR. Consequently, this model is also a robust focal plane-based wavefront error estimator.

%\subsubsection{$\beta$ and $\alpha$ estimates}

\subsubsection{Secondary parameters estimates}
\label{SS:width}

We have emphasized in Sect. \ref{S:th} that parameters $\alpha$ and $\beta$ govern the PSD shape and assist in the AO correction diagnosis.

\begin{figure*}[h!]
    \centering
    \includegraphics[width=0.47\linewidth]{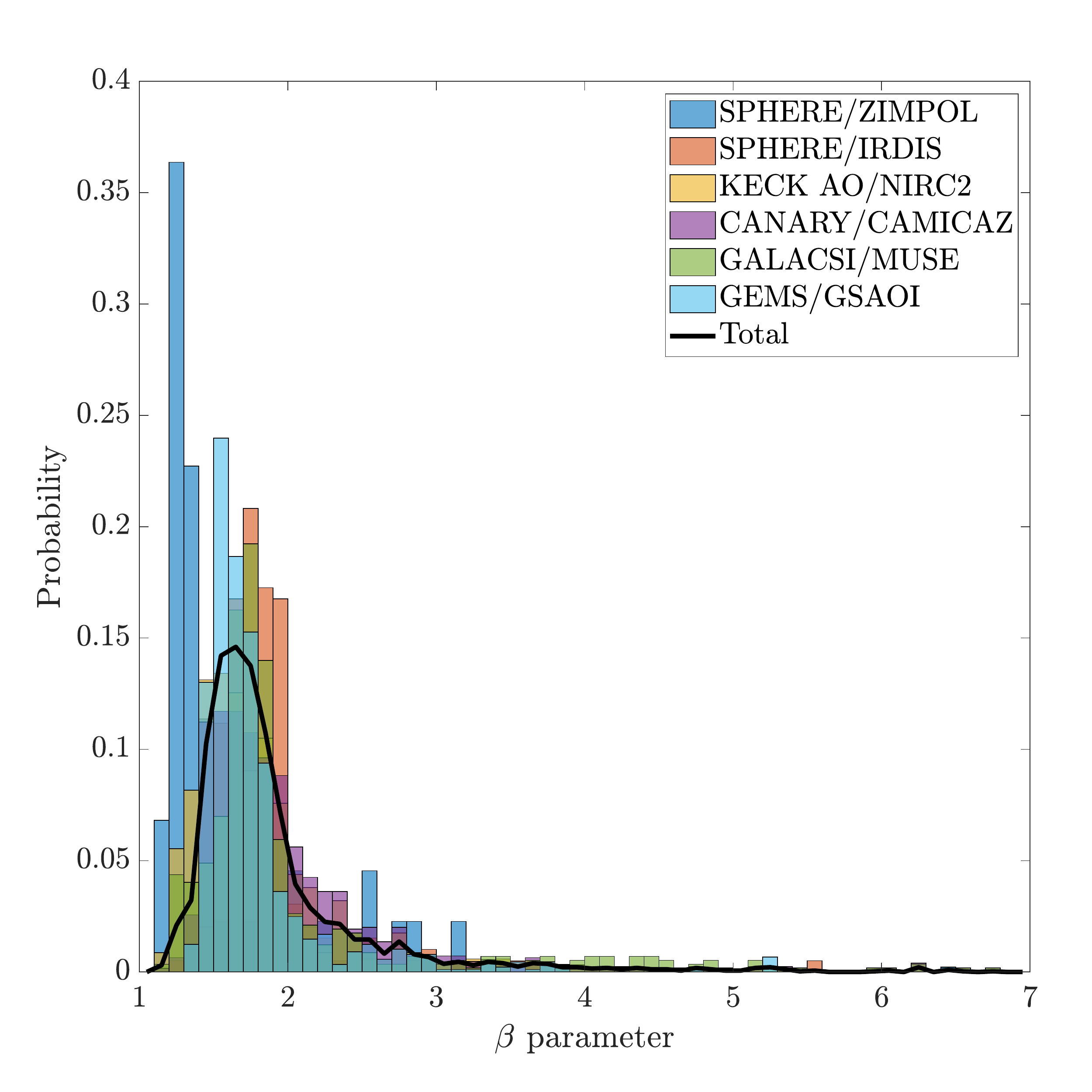}
    \hspace{0.02\linewidth}
     \includegraphics[width=0.47\linewidth]{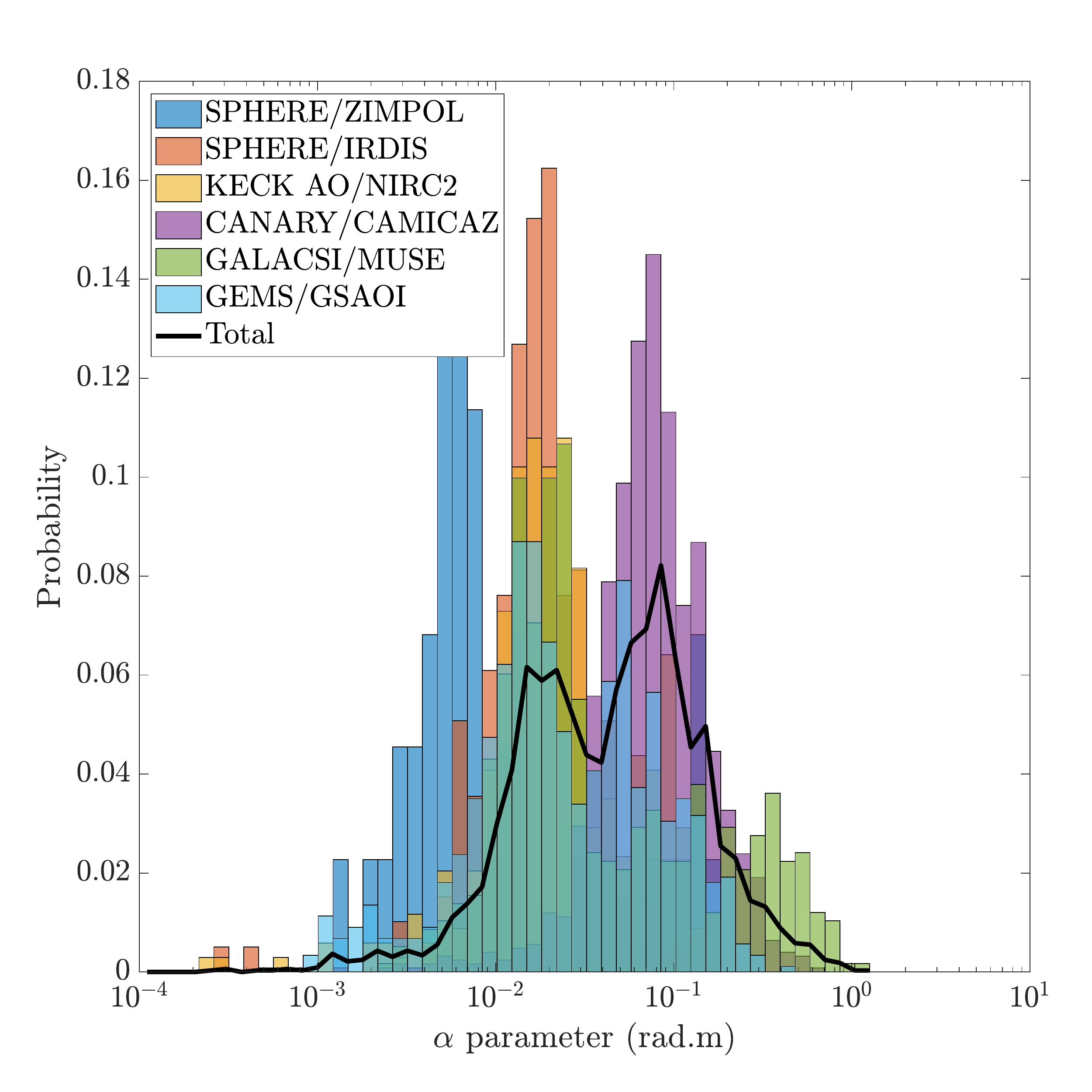}
    \caption{Histograms for the $\beta$ and $\alpha$ parameters for each system, as well as the averaged distribution.}
    \label{fig:beta}
\end{figure*}

Figure \ref{fig:beta} presents histograms for each system; the $\beta$ parameter has a relatively narrow histogram with a peak located at $\beta = 1.82$ and a 1-$\sigma$ standard deviation of 0.6, while the $\alpha$ histograms seem particularly instrument-dependent with values from 0.001\,rad.m for efficient AO correction to 1\,rad.m for marginal AO correction (e.g., in the bluest visible wavelengths of MUSE). Our first conclusion here is that an optimized AO system should provide a PSD with a $\beta$ parameter comprised between 1 and 1.9, as discussed in Sect. \ref{S:th}. For larger values of $\beta$, there is an excess of residual error into low-order spatial frequencies. For instance, we see that on Keck AO/NIRC2 images the median $\beta$ increases from 1.7 up to 2.1 in NGS and LGS modes, respectively, indicating the presence of additional low-order modes introduced here mainly from the focal anisoplanatism \citep{Wizinowich2012,VanDam2006}. In addition, we have observed cases with $\beta>2.5$ on Keck data due to a strong wind-shake effect that was enlarging the PSF FWHM by a factor three compared to nominal performance. Both $\alpha$ and $\beta$ parameters grow up in the presence of strong wind shake and are wavelength dependent as we see on GALACSI/MUSE histograms. However, from GEMS/GSAOI data analysis, we do not find clear trend with respect to the field position, which may be owing to the uniform correction across the field provided by the MCAO mode of GEMS.

We illustrate here that those two parameters carry additional and relevant information on the AO correction. However, the exact connection with the AO status is not straightforward to identify. \ccc{To enable this identification, we are currently developing a convolutionnal neural network \citep{Herbel2018} that we train to estimate the model parameters from the AO control loop data, such as wavefront measurements.} 
As we can collect a very large amount of data for the purpose of estimating a few tens of parameters, solving this problem is definitely achievable with dedicated simulations and on-sky data that we will continue to be collected regularly among observatories.

\subsection{Discussion on the influence of exposure time and spectral bandwidth}

\begin{figure*}[h!]
    \centering
    \includegraphics[width=0.47\linewidth]{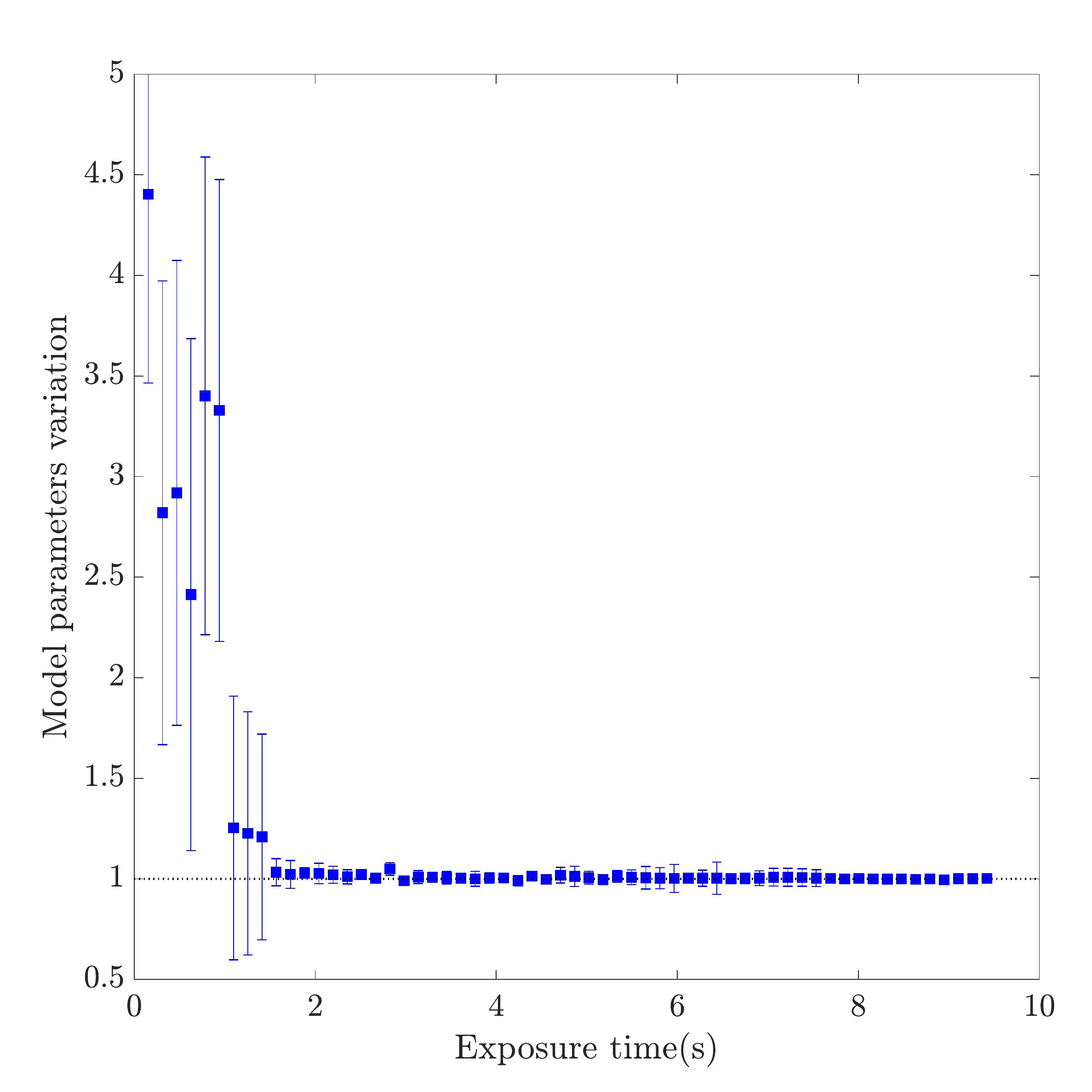}
    \hspace{0.02\linewidth}
    \includegraphics[width=0.47\linewidth]{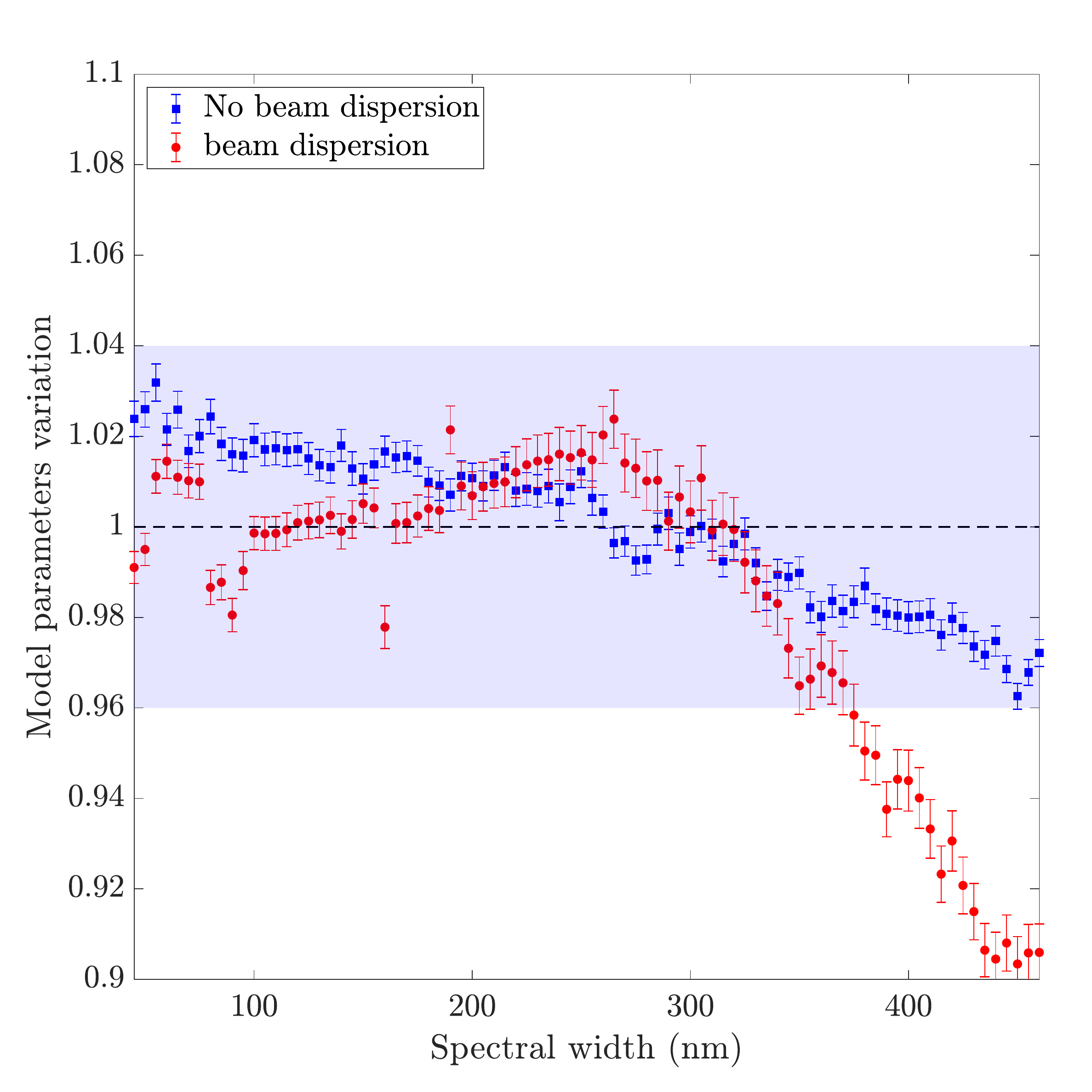}
    \caption{\textbf{Left:} Average of absolute retrieved parameters with respect to the exposure time. The parameters are normalized by the value obtained from the long-exposure PSF (60 frames). Error bars are assessed by averaging results over the SOUL/LUCI data sets. \textbf{Right:} Average retrieved parameters normalized by the median value over the whole span with respect to the spectral width. The envelope shows the $\pm$ 4\% of variations around the median value and the spectral bandwidth is systematically centered around 700\,nm. Error bars are assessed by averaging results over the GALACSI/MUSE data sets.}
    \label{fig:exp}
\end{figure*}

One of the major assumptions in the model proposed by \citet{Fetick2019_Moffat_aa} concerns the infinitely long exposure hypothesis. This model is therefore not capable of reproducing atmospheric speckles that average when taking a sufficiently long exposure. As the method relies on the second-order statistical moment of the residual phase, the time necessary to achieve a convergence of the PSD shape is highly dependent on atmospheric parameters but may be achieved in few seconds \citep{Martin2012}. Thanks to SOUL/LUCI data, we analyzed the PSF accuracy when fitted on short exposure images with integration times from 0.157\,s to 60\,s. We find that the model matches the PSF down to 1-2\,s of exposure \ccc{and for a PSF acquired at 1.6\,$\mu m$}, \ccc{below which the parameters estimation begins to degenerate as illustrated in Fig. \ref{fig:exp} (left) through the average of absolute parameters variations. We have also noticed that the PSF shape remains stable from few seconds exposure, which explains the stability of retrieved parameters.}

Moreover, using GALACSI/MUSE data, we tested the model on a polychromatic image, from 2.5\,nm up to 470\,nm of spectral width by binning monochromatic PSF observed simultaneously with MUSE. When compensating for the beam dispersion (recentering PSFs and then stack), the parameter estimation does not deviate by more than 4\% over the whole spectral width span as presented in Fig. \ref{fig:exp} (right). Consequently, the presence of chromatic static aberrations and atmosphere chromaticity do not prevent the model from  very accurate characterization of the PSF. When including the beam dispersion (no recentering before stacking), which is 350\,mas (14 pixels) for the largest spectral width, the parameter estimation degrades up to 10\% and goes beyond the 4\% threshold after 300\,nm of spectral width. As a result, the model remains robust and reliable, even in the presence of such a strong beam dispersion. 

\section{Joint atmospheric, AO performance and static aberrations retrieval}
\label{S:static}

The previous section illustrates that this model  accurately and reliably reproduces the AO residual component of the PSF. Considering that the model is parsimonious (7 AO parameters and 4 \ccc{photometric and astrometric parameters}) and the large amount of pixels (10,000 for a 100$\times$100 image) that the model-fitting may rely on, one can attempt to retrieve more degrees of freedom, such as static aberrations.

\ccc{In order to estimate the static coefficients, we replace in the criterion given in Eq. \ref{E:crit} $\otf{\text{Static}}$ by $\otf{\text{Static}}(\boldsymbol{\mu}_\text{Static})$, which becomes a parametric OTF term. Besides, determining the wavefront from a single long-exposure image suffer from strong degeneracies by essence and this approach does not overcome this problem by using a non-linear minimization algorithm to estimate the problem solution. We distinguish two sorts of degeneracy
\begin{itemize}
    \item Sign ambiguity : this occurs when trying to fit pair modes for which the PSF is not sensitive to the sign \citep{Mugnier2008}.  Therefore, we concentrate the static aberrations retrieval on piston, tip and tilt modes only. Moreover,  we bound the solution domain between -$\lambda/2$ to $\lambda/2$ in order to overcome phase wrapping issues with piston modes.
    \item Local minima : two different wavefront patterns may produce a similar but slightly different PSF and make the PSF-fitting algorithm converge towards a local minimum of the criterion. To assess how much this problem affect our algorithm, we have compared the PSF model with various conditions of piston, tip and tilt levels in order to highlight possible combinations (10,000) of modes that could produce a similar PSF. From a vector of aberrations, we have tested different permutations of the elements of this vector and in 99\% of cases, the relative PSF variation is larger than 1\%, which is large enough to change the structure of the PSF and retrieve the correct wavefront map, as long as the S/N is sufficient (>50). We also rely on an analysis from \citet{Gerwe2008} that shows that for a fit of 35 Zernike polynomials on a segmented pupil, there are no local minima as long as the initial guess for the static coefficient remains within $\pm$ 0.2$\lambda$ = 330 nm rms in H-band from the optimal solution. We are in a situation where the static aberrations we attempt to retrieve are already mitigated from the Keck telescope active control \citep{Chanan1988} and the VLT spiders coating \citep{Milli2018_LWE}. Consequently, the aberrations level we must retrieve remains sufficiently weak to mitigate the presence of local minima. At a level greater than 300\,nm (Keck AO residual is 280\,nm) would impact drastically the PSF and degrade the telescope science exploitation so much that this aberration would be necessarily visible and mitigated as much as possible.
\end{itemize}
Henceforth, we are in a good situation to estimate piston, tip and tilt static modes on Keck and VLT images obtained on bright star.}

\subsection{Keck cophasing error retrieval}
\label{SS:piston}

The analysis presented in this section focuses on a smaller sample of 129 Keck AO/NIRC2 data acquired in NGS mode and with a high SR (>40\%) in 2013 \citep{Ragland2016}. The residual NCPA map was calibrated at the beginning of the night. We compared model-fitting performance in three situations: (i) a fit of PSD and photometric and astrometric parameters (11 values), (ii) a fit of these parameters when including, following Eq. \ref{E:otfStatic},  the static aberration map calibrated at the beginning of the night and, (iii) a joint adjustment of the PSD, photometry/astrometry and the 36 piston values corresponding to each Keck pupil segment (47 parameters in total), including the calibration static aberration map.

Figure \ref{fig:psfStat} shows the PSFs and the residual maps produced from the three model-fitting strategies considered in this section and obtained for three cases acquired in February, August, and September, 2013. We firstly conclude that accounting for the residual NCPA static aberrations improves the model fitting as revealed by the residual map. This is also confirmed by the results shown in Tab. \ref{tab:srkeck} which gives the median and $1\-\sigma$ standard deviation of SR/FWHM estimates over the 129 data samples as well as the mean square error (MSE) determined from the residual map. \ccc{We firstly observe that biases and dispersion values on the SR are larger than in Tab. \ref{tab:statSR}, owing to the data sample we consider in the present analysis and that gathers only high SR data for which the residual atmospheric aberrations are weaker. This advocates for including the calibrated static aberrations map into the model in order to mitigate this bias in the model.  From Fig. \ref{fig:psfStat}, we also observe that these static aberrations manifest mainly as static speckles in the PSF and consequently contribute marginally to shape the PSF core, which explain why the statistics on the FWHM estimates are not influenced by the reduction of the data sample.} Moreover, the results shown in both Fig. \ref{fig:psfStat} and Table \ref{tab:srkeck} highlight that the model accuracy is slightly improved by including the NCPA residual map in the model, with still 50\% improvement on the MSE. Furthermore, the cophasing map retrieval allows to drastically enhance the model-fitting results, for example by up to a factor three on the MSE compared to the first scenario, as also illustrated in Fig. \ref{fig:psfStat} and Tab. \ref{tab:srkeck}. The fact that the model efficiently reproduces the speckles observed on sky images suggests that these speckles are mainly introduced by cophasing errors due to the telescope.  

\begin{table}[h!]
    \centering
    \begin{tabular}{|c|c|c||c|c||c|}
    \hline
    Strategy &\multicolumn{2}{c||}{ $\Delta$ SR (pts)} & \multicolumn{2}{c||}{ $\Delta$ FWHM (mas)} & MSE (\%)  \\
    \hline
    (SR>40\%)& Median & std & Median & std  &   \\ 
    \hline
    (i) & 1.4 & 1.9 & -0.9 & 0.8 & 0.81\\
    \hline
    (ii) & 1.0 & 1.8 & -0.7 & 0.8 & 0.54\\
    \hline
    (iii) & 0.7 & 1.1 & -0.5 & 0.4 & 0.32\\
    \hline
    \end{tabular}
    \caption{Median and 1-$\sigma$ dispersion of SR and FWHM errors and mean square error obtained over the 129 H-band and high SR (>40\%) Keck/NIRC2 images treated in this analysis for the three implemented strategies: (i) PSD/stellar parameters retrieval, (ii) including the residual NCPA map in the model, and (iii) also including the estimation of Keck II cophasing errors. Systematically, accounting for the NCPA map allows us to slightly reduce the biases on the estimates  and diminish the MSE, but best model-fitting results are obtained thanks to cophasing error retrieval that decreases the MSE by almost a factor three.}
    \label{tab:srkeck}
\end{table}

\begin{table*}[h!]
    \centering
    \begin{tabular}{||c||cc||cc||cc||c||}
     \hline
    \hline
    & \multicolumn{2}{c||}{Case 1 - February 2013} & \multicolumn{2}{c||}{Case 2 - August 2013} & \multicolumn{2}{c||}{Case 3 - September 2013} & \makecell[bb]{ \multirow{5}{*}{ \includegraphics[width=0.047\linewidth]{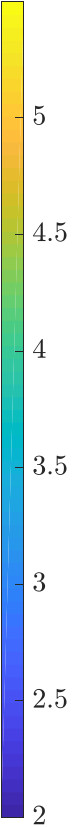}} \vspace{-.7cm}}\\
   \hline
    \hline
   
    \makecell[bc]{Sky \\ image \vspace{1cm}} & \multicolumn{2}{c||}{\includegraphics[width=0.2\linewidth]{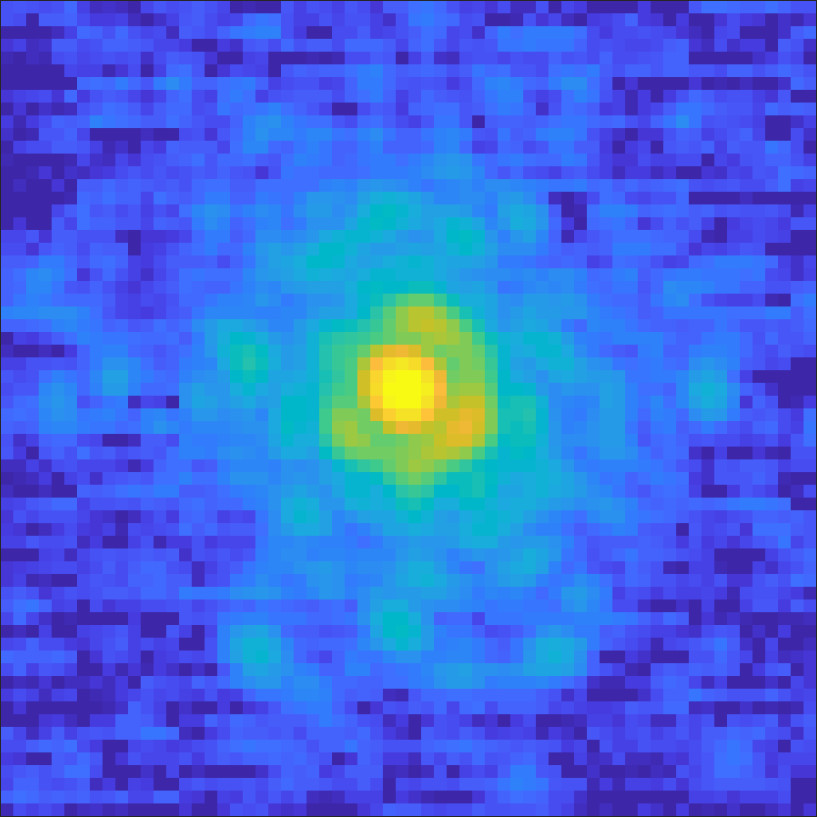}} & \multicolumn{2}{c||}{\includegraphics[width=0.2\linewidth]{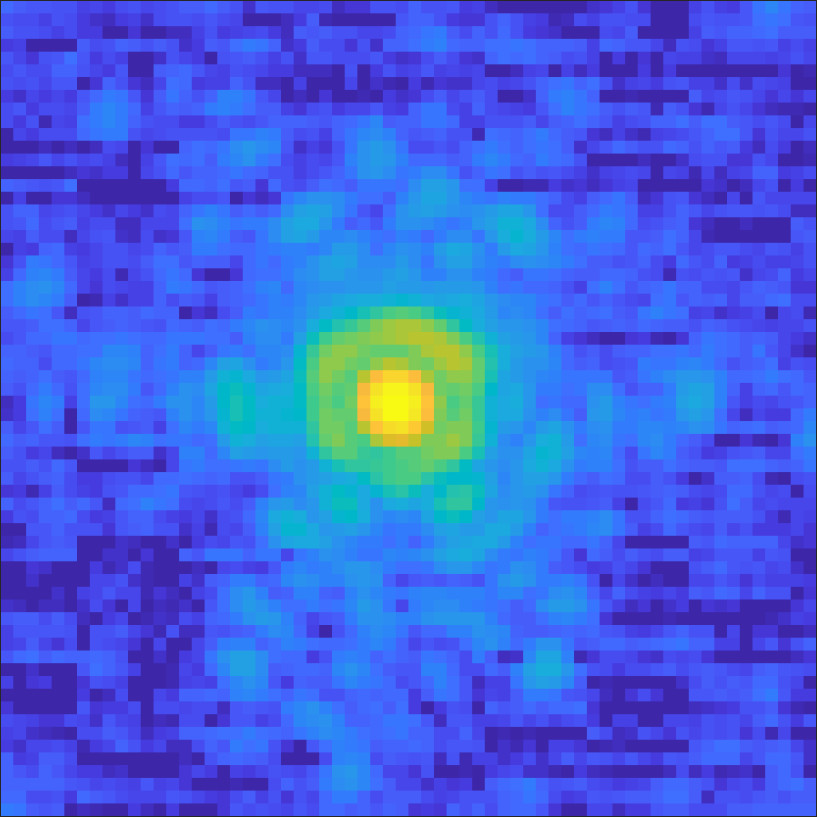}}  & \multicolumn{2}{c||}{\includegraphics[width=0.2\linewidth]{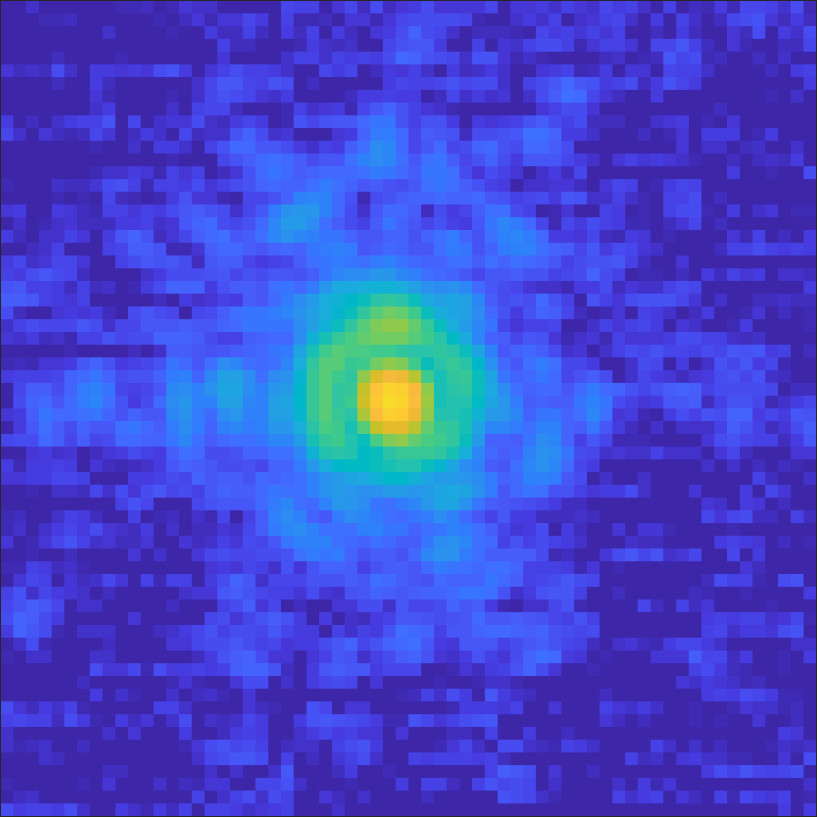}} & \\

    \makecell[bc]{Model \vspace{0.8cm}}& \includegraphics[width=0.1\linewidth]{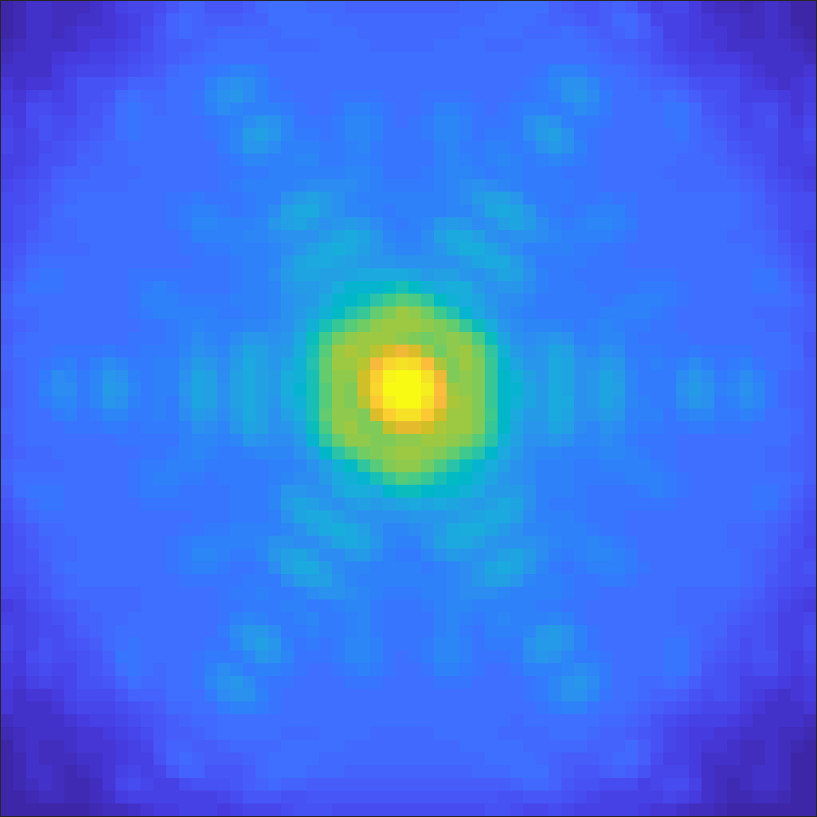} & \includegraphics[width=0.1\linewidth]{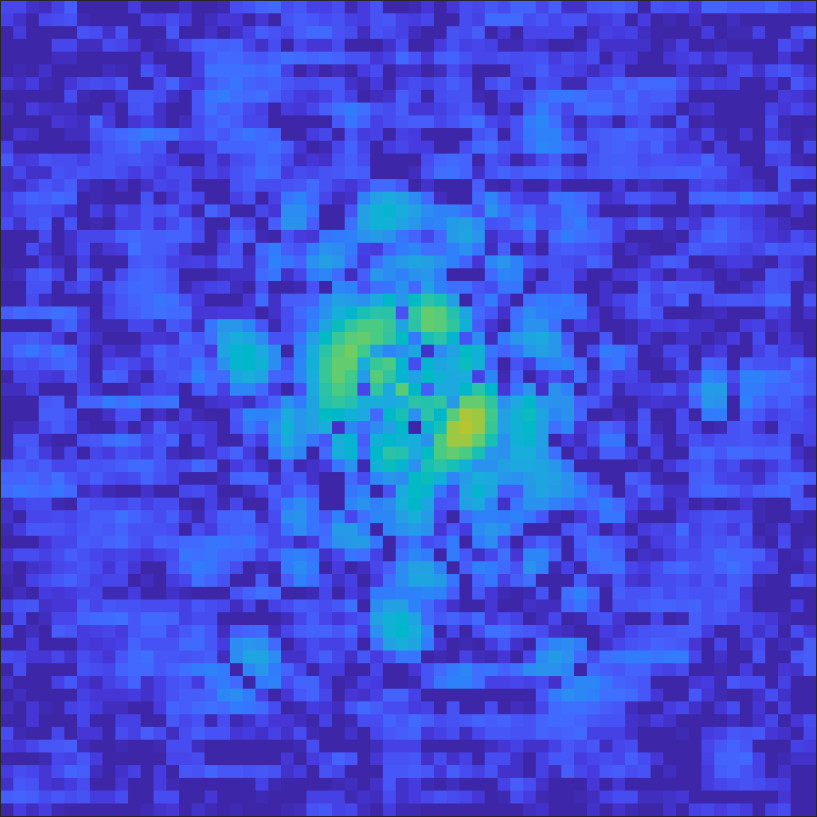}&\includegraphics[width=0.1\linewidth]{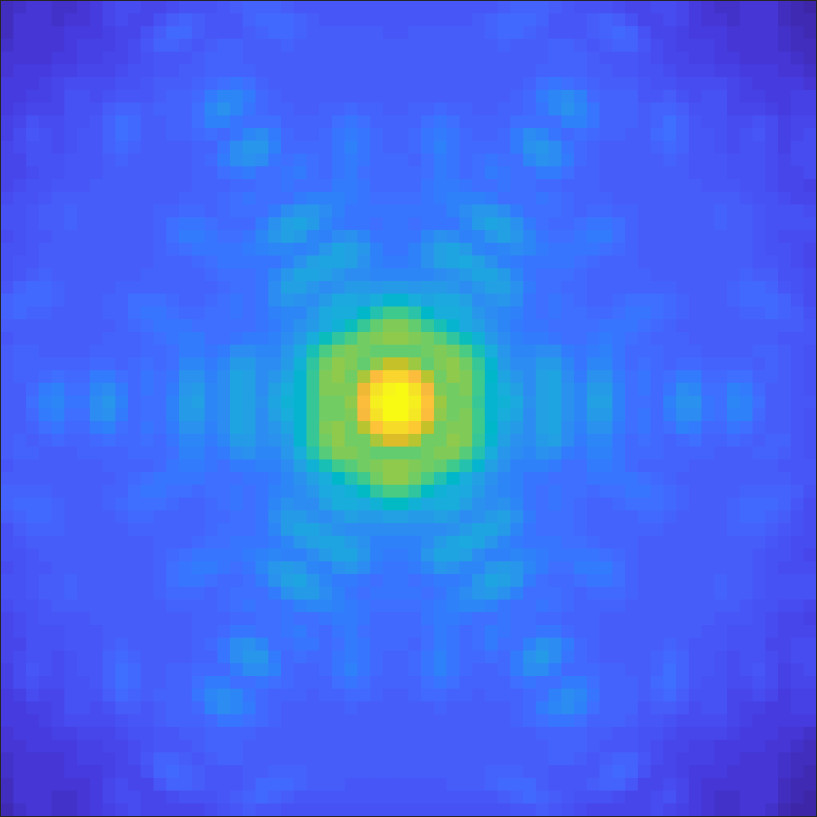}& \includegraphics[width=0.1\linewidth]{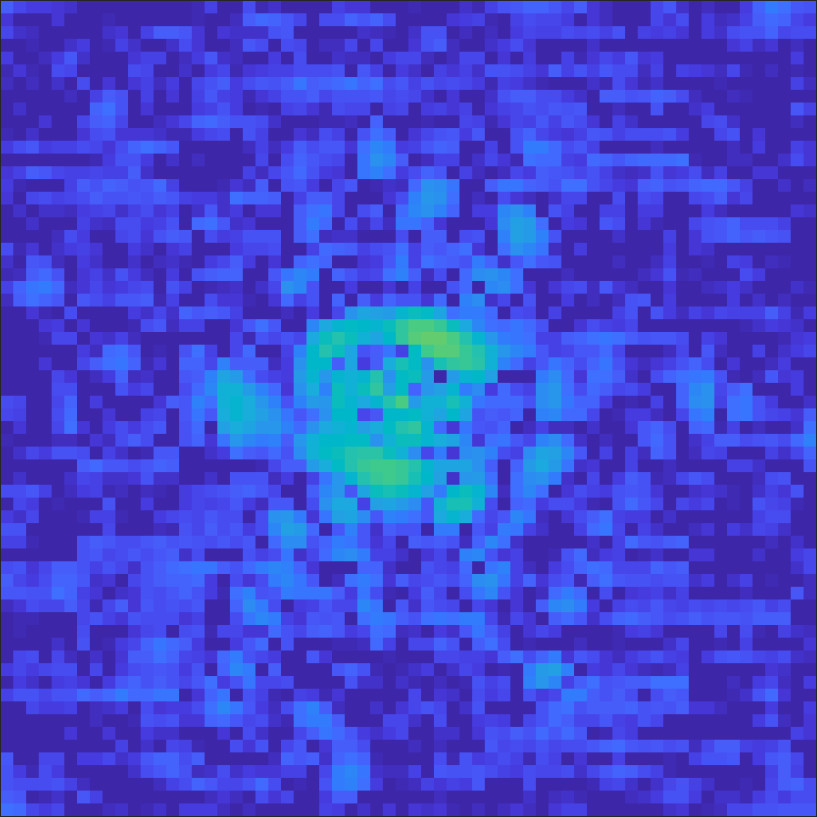}& \includegraphics[width=0.1\linewidth]{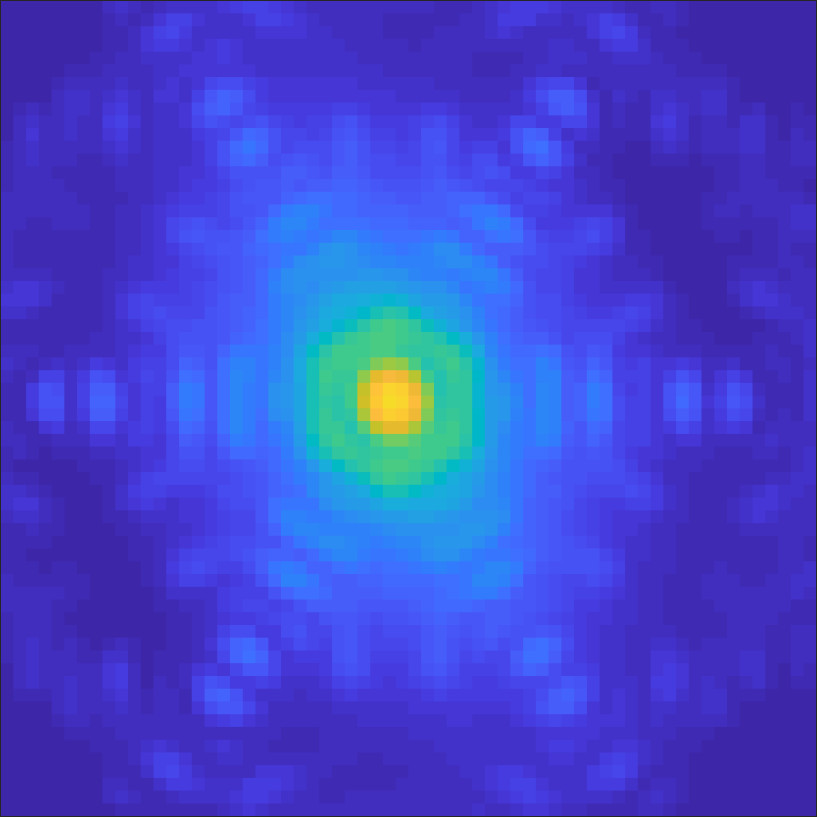}& \includegraphics[width=0.1\linewidth]{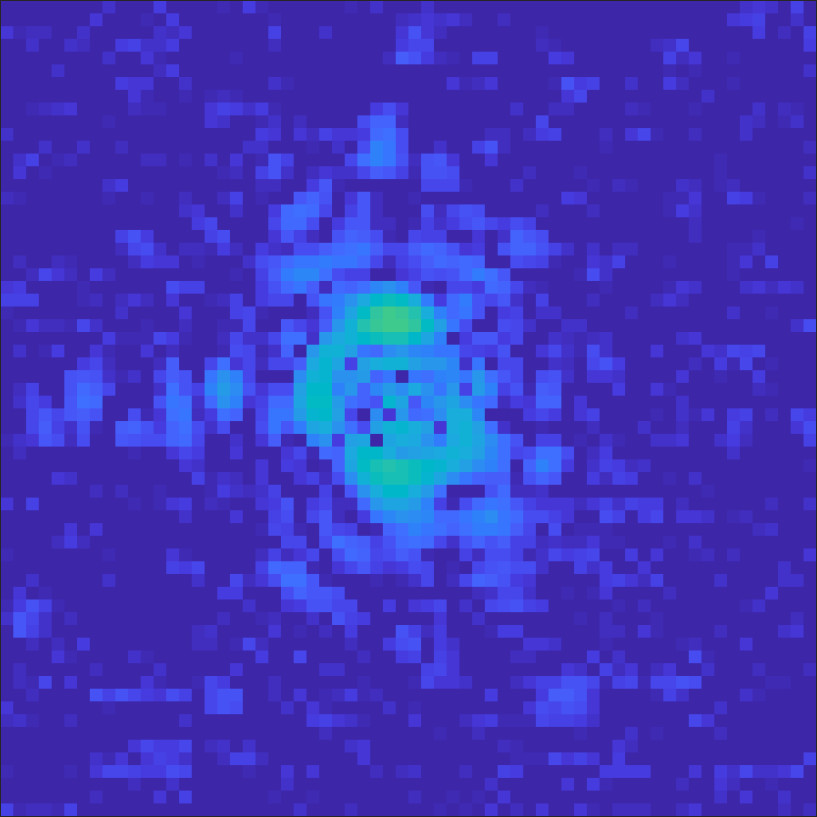} &\\

    \makecell[bc]{Model \\ +\\ NCPA \vspace{0.3cm}}& \includegraphics[width=0.1\linewidth]{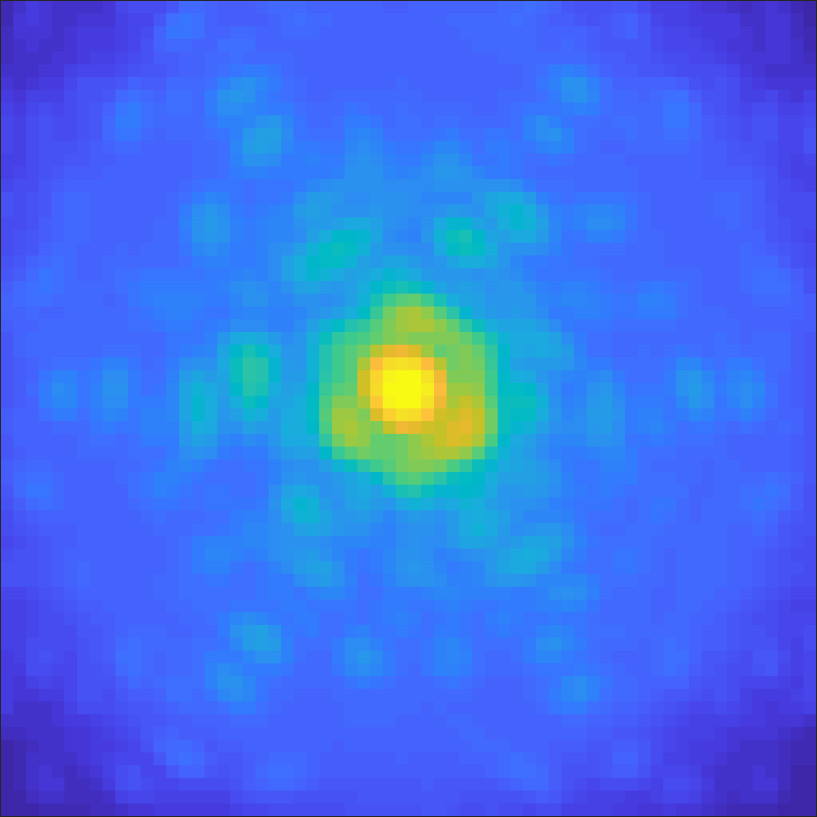} & \includegraphics[width=0.1\linewidth]{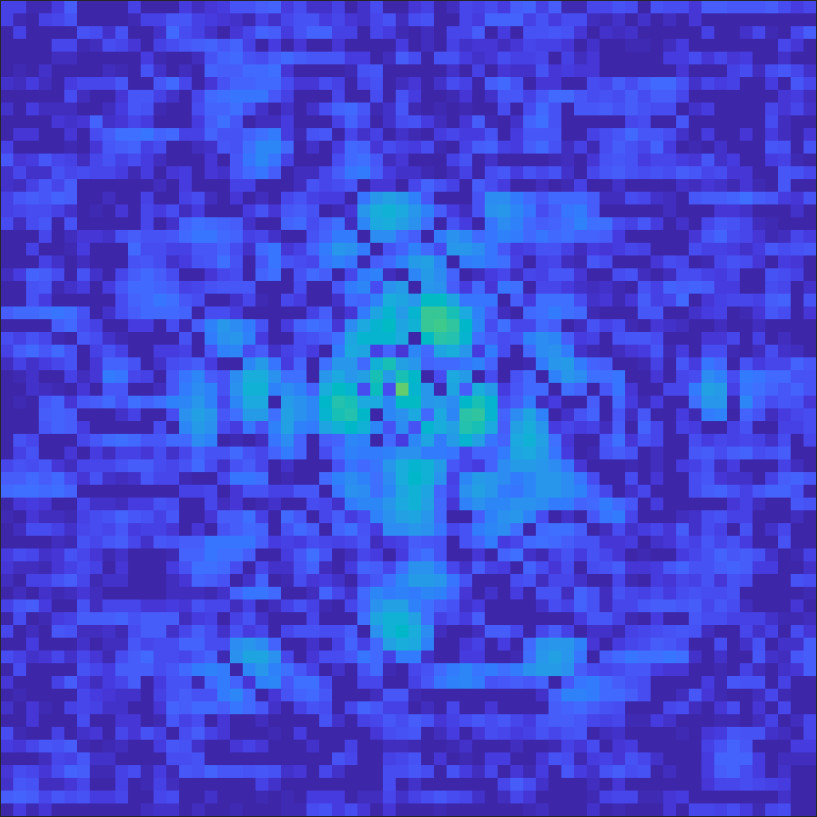}&\includegraphics[width=0.1\linewidth]{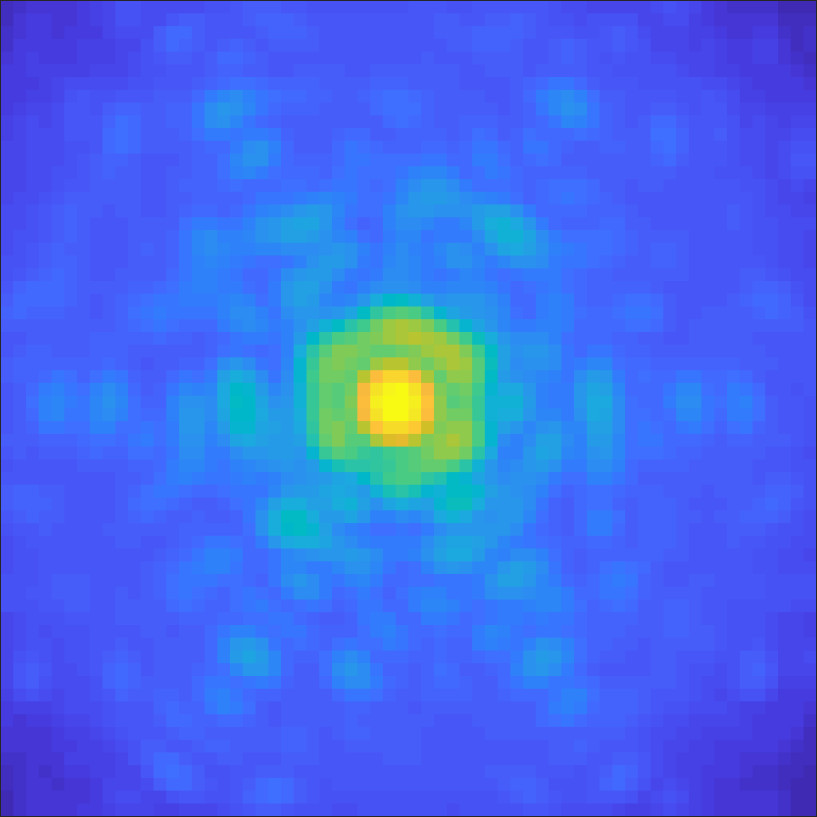}& \includegraphics[width=0.1\linewidth]{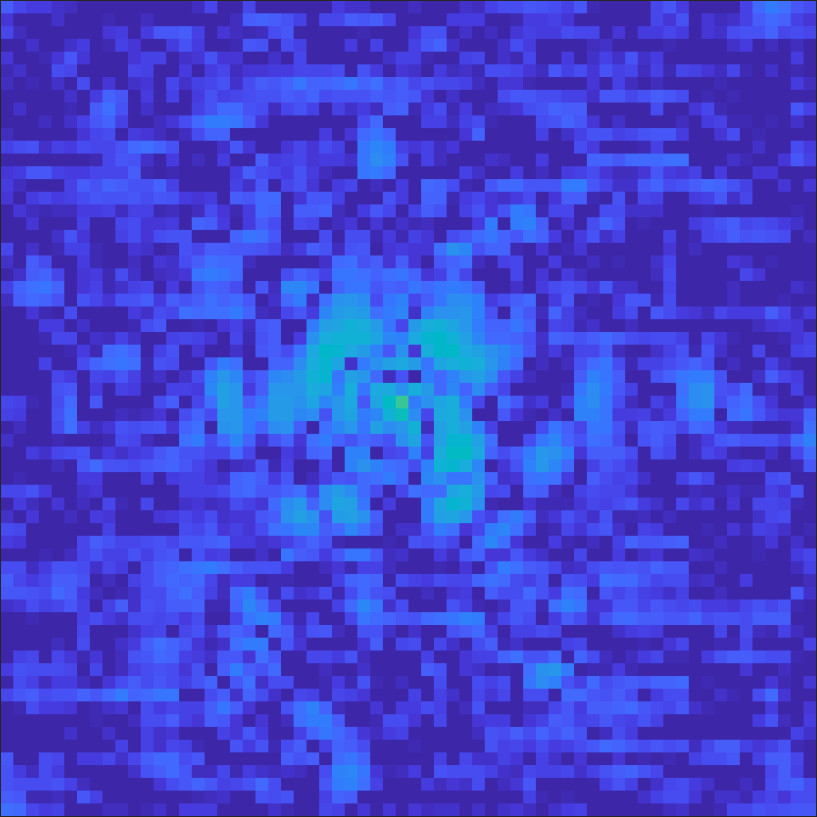}& \includegraphics[width=0.1\linewidth]{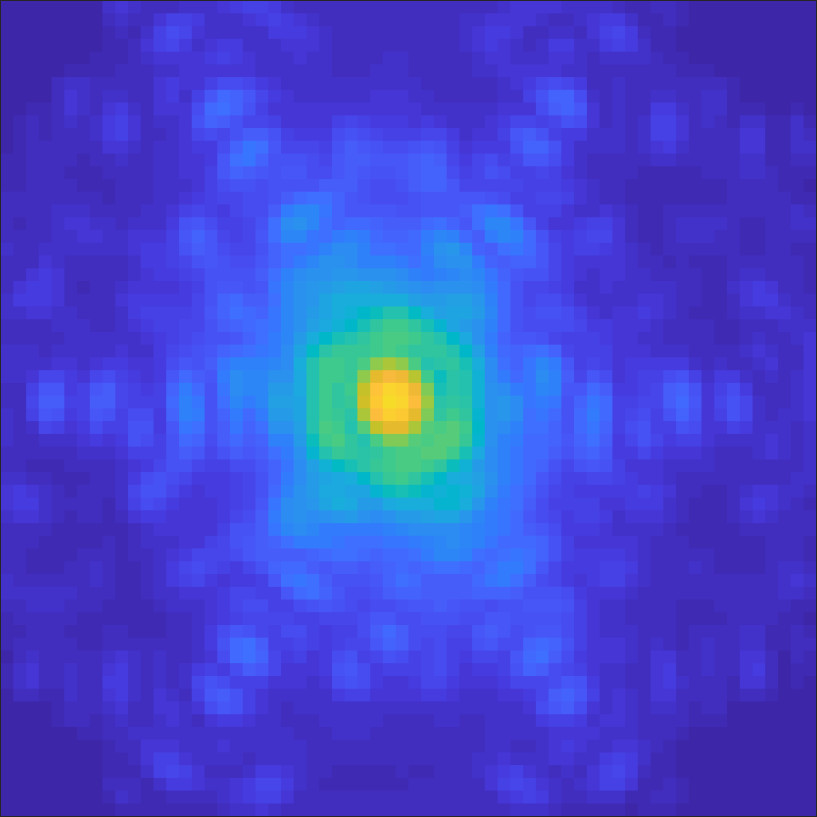}& \includegraphics[width=0.1\linewidth]{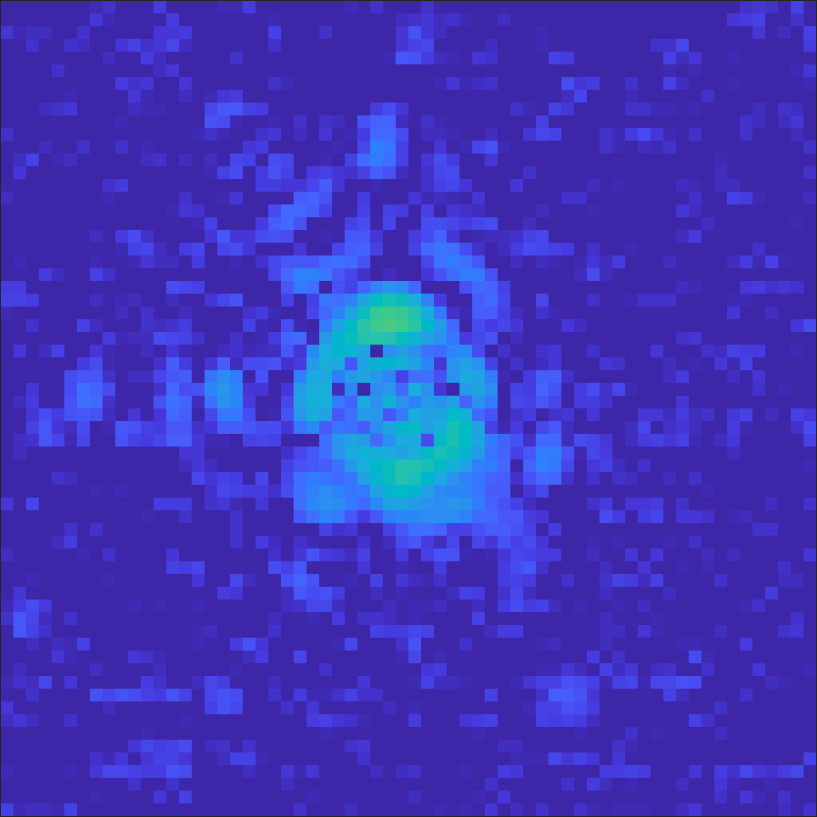} &\\

    \makecell[bc]{Model \\ +\\ NCPA \\ +\\ Cophasing \vspace{0.01cm}} & \includegraphics[width=0.1\linewidth]{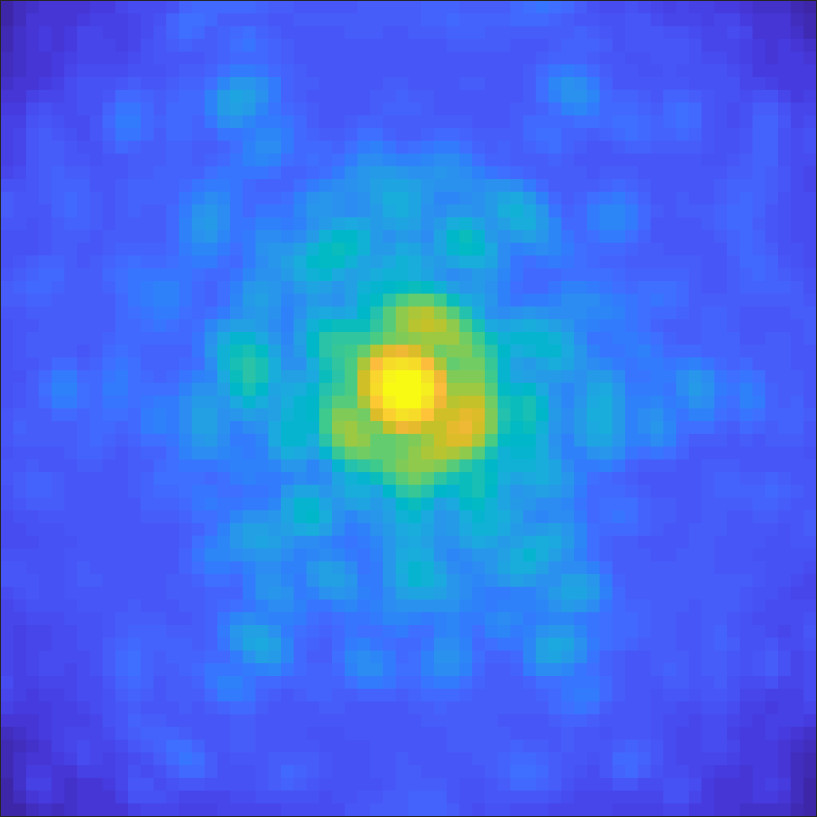} & \includegraphics[width=0.1\linewidth]{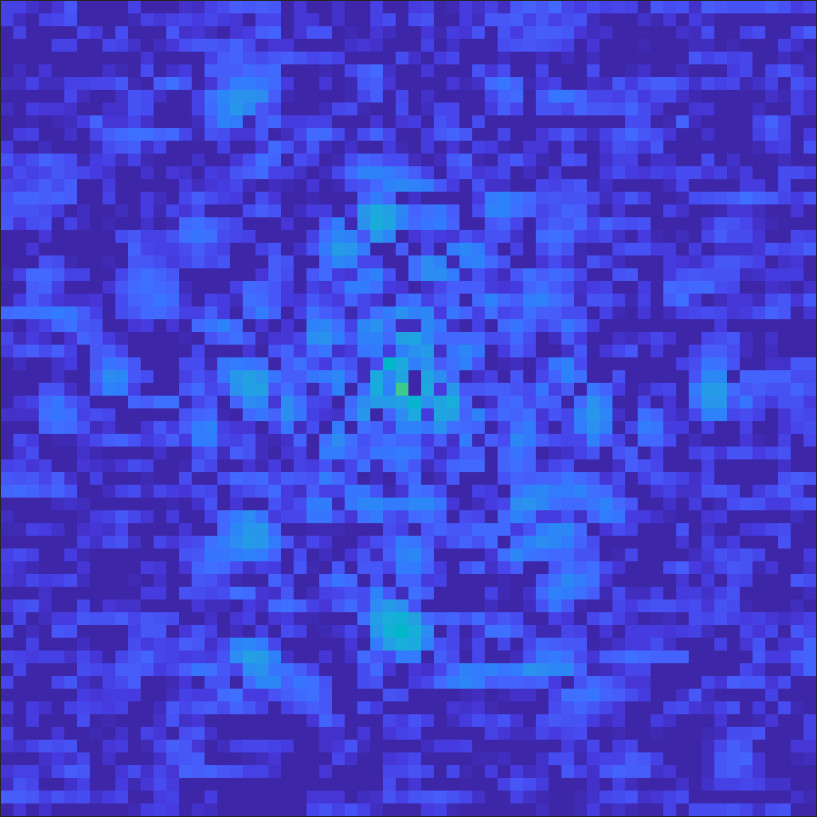}&\includegraphics[width=0.1\linewidth]{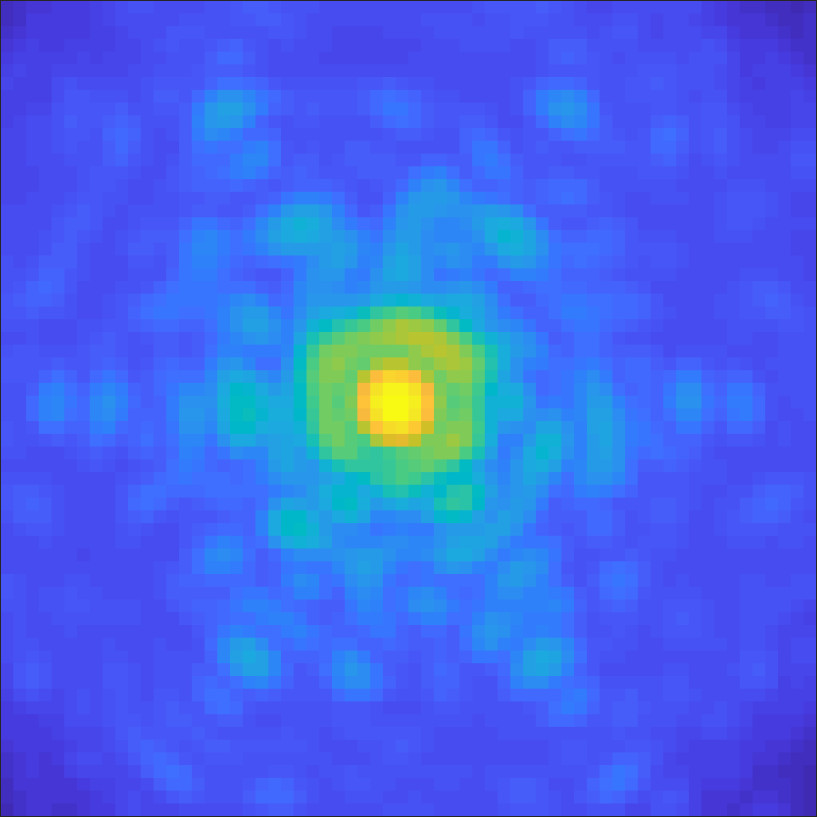}& \includegraphics[width=0.1\linewidth]{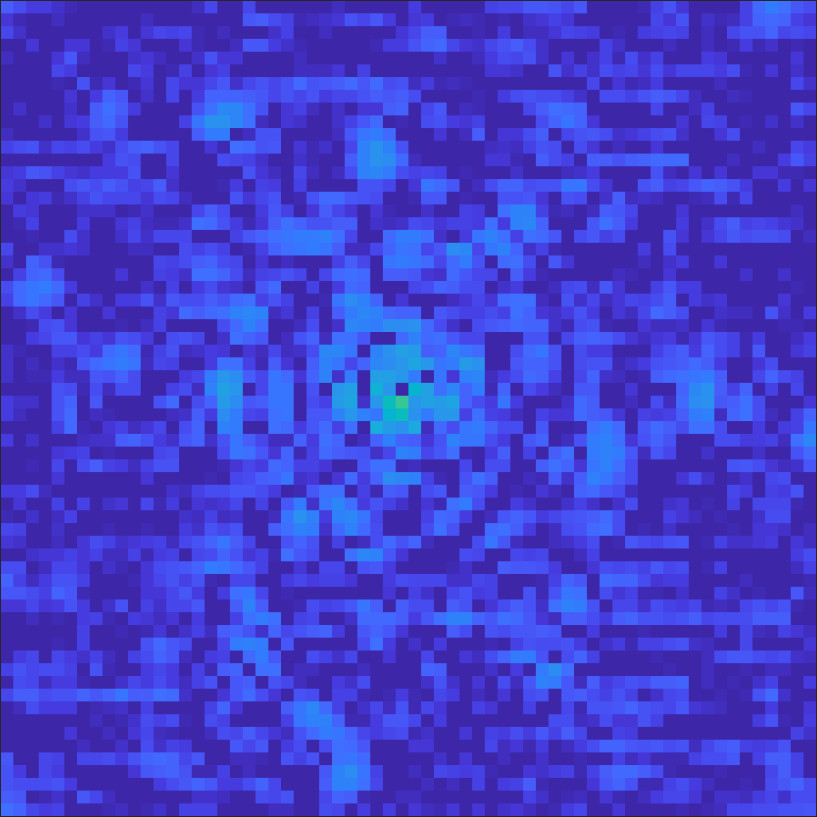}& \includegraphics[width=0.1\linewidth]{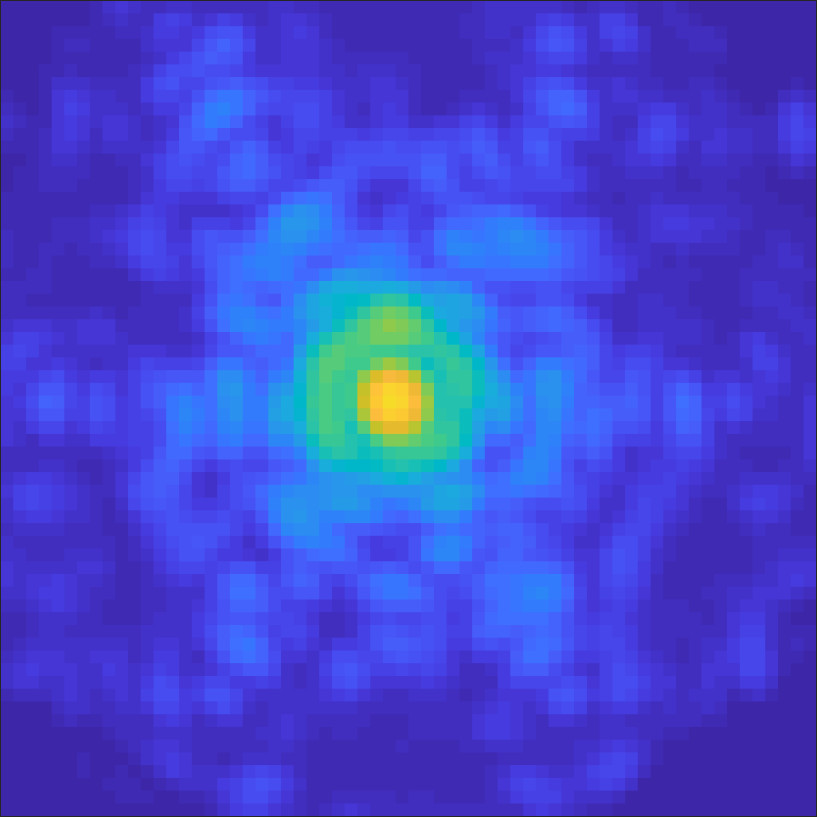}& \includegraphics[width=0.1\linewidth]{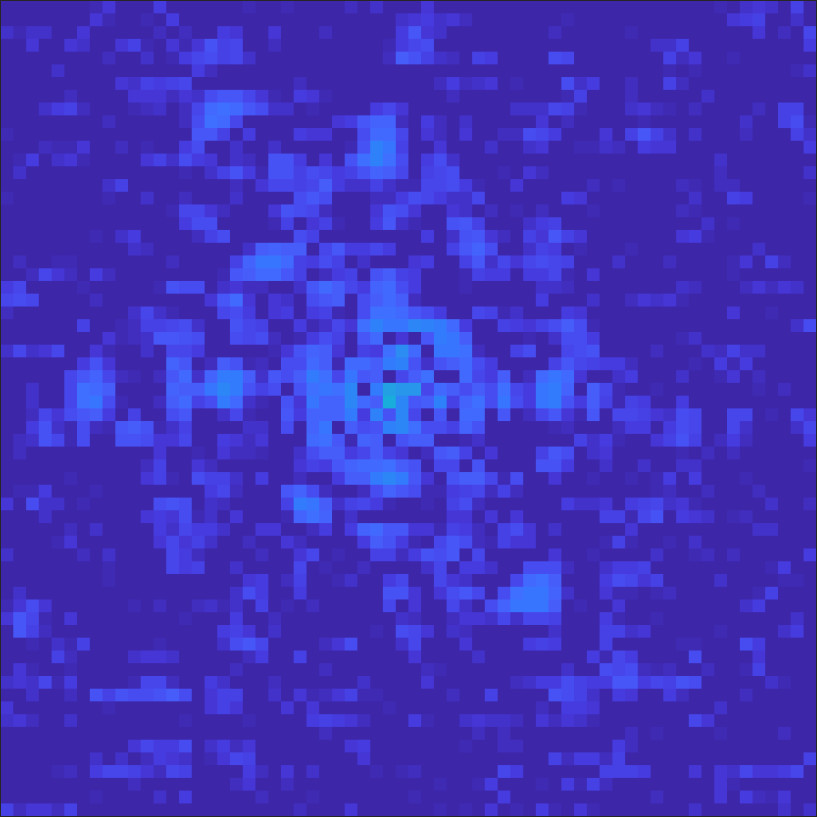} &\\
    \hline
    \hline
    
    \end{tabular}
    \captionof{figure}{Comparison in log scale of on-sky PSFs and fitted  model using the three implemented strategies. Each column correspond to a single data sample acquired at different epochs. For each model configuration, we report both the adjusted PSF and the residual map. This figure illustrates that the cophasing map retrieval improves drastically and consistently the residual on top of the gain brought by including the NCPA residual map.}
    \label{fig:psfStat}
\end{table*}

\begin{figure*}[h!]
    \centering
    \includegraphics[width=\linewidth]{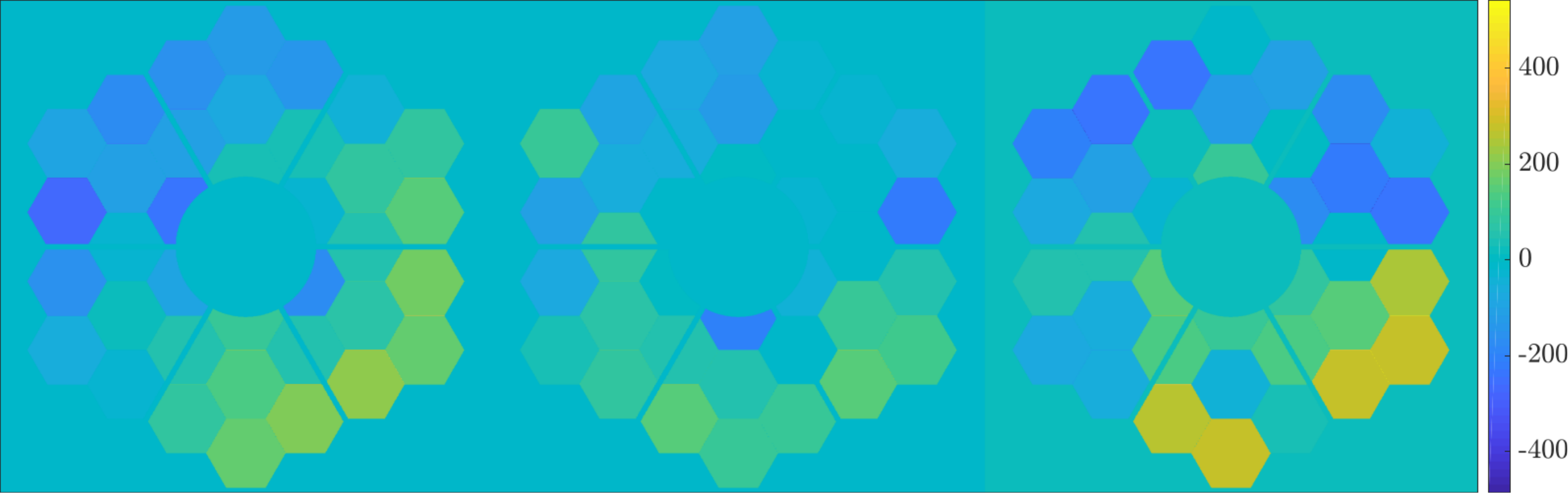}
    \caption{Retrieved cophasing maps in nm from PSFs acquired during three different epochs in 2013. The telescope elevation and azimuth were respectively (79,240), (85,-82) and (43,240) from left to right.}
    \label{fig:statmap}
\end{figure*}

In addition, in order to further demonstrate the strength of this framework, we present in Fig. \ref{fig:statmap} the retrieved cophasing maps for three specific cases. The corresponding NCPA map used in the model is presented in \citet{Ragland2016}. We unmistakably observe the presence of a stair mode, \ccc{that introduces a constant step between adjacent segment rows or columns} and that has been already highlighted from data acquired in 2017 using an alternative technique \citep{Ragland2018_COPHASING}.

Moreover, each of those three images was obtained with different elevation-azimuth telescope positions that are reported in Table \ref{tab:statStd}, which shows that the wavefront standard deviation value is very close to observations reported by \citet{Ragland2018_COPHASING} for a 45$^\circ$ telescope elevation. Also, the stair case peak-to-valley (PV) energy increases with low telescope elevation and rotates with respect to the telescope azimuth. This observation reveals a connection between the presence of this stair mode and global flexure over the primary mirror that is controlled to compensate for the gravity effect. Deeper analyses will confirm the presence of such a residual flexure on the Keck telescoped and whether or not this framework now to allow us to correct for this stair mode to generally improve the scientific exploitation of the Keck, but also future segmented telescopes.

\begin{table}[]
    \centering
    \begin{tabular}{|c|c|c|c|}
    \hline
    Case & Elevation/Azimuth ($^{\circ}$) & PV (nm) & std (nm) \\
    \hline
    Feb. 2013 & (79,240)  & 760 & 95\\
    \hline
    Aug. 2013 & (85,-82)& 530 & 66\\
    \hline
    Sept. 2013 & (43,240) & 1030 & 118 \\
    \hline
    \end{tabular}
    \caption{Summary of peak-to-valley (PV) and 1-$\sigma$ standard deviation of retrieved cophasing error map regarding the corresponding telescope elevation/azimuth.}
    \label{tab:statStd}
\end{table}

\ccc{Measuring the piston map from the focal-plane image is feasible from results obtained with the present analysis, but necessitates to have a star bright enough in the field to calibrate these aberrations}. We must ensure that the shape of the primary mirror is controlled without the aid of a focal-plane-based technique in order to achieve the best image quality regardless of the observed field. We are again in a situation where we have to retrieve a few parameters (47) from a potentially large amount of data delivered by segment position sensors, temperature, pressure and humidity sensors, atmospheric parameters near the dome and even AO telemetry, which can deliver information about the piston error when using a pyramid WFS \citep{Bond2018_pyramid,Schwartz2018}. It is not easy to investigate the connection between this ensemble of data and the piston map, and the use of neural networks for this purpose will be explored.

\subsection{SPHERE low wind effect retrieval}
\label{SS:lwe}

the analysis presented in this section focuses on 176 SPHERE/IRDIS data obtained with good SR conditions (>40\%) so as to test two different fitting strategies, namely (i) fitting the PSD and photometry/astrometry parameters (11 parameters) and (ii) fitting these latter 11, and 12 additional parameters corresponding to piston, tip, and tilt of each VLT pupil area delimited by the spiders in order to account for the LWE. According to \citet{Sauvage2016}, this description of the LWE allows the main impact of this effect to be reproduced, i.e., a strong PSF asymmetry.

Figure \ref{fig:psf_lwe} presents the results of the two strategies implemented in our study over three particular cases for which a strong LWE is observed. We observe, especially in Fig. \ref{fig:psf_lwe}, that the sole parametrization of the PSD is not rich enough to precisely reproduce  the strong asymmetry, which becomes a solved problem thanks to the aberration parametrization we propose.  Regarding the estimated PSF shape and values given in Table \ref{tab:srsphere}, we show clear evidence that (i)  describing the LWE as a combination of differential piston, tip, and tilt allows to accurately characterize the PSF asymmetry, and (ii) accounting for this description in the PSF model ensures that biases and dispersion on SR and FWHM estimates are drastically mitigated. Moreover, the average MSE value over the 176 data samples is significantly diminished, that is, by a factor two. \ccc{ On the three cases illustrated in Fig. \ref{fig:psf_lwe},Table \ref{tab:case_lwe} also reports the SR and FWHM errors as well as the MSE obtained from the PSF-fitting. These results show clearly a gain on accuracy for these particular cases for which the LWE impact on the PSF is significant comparatively to the statistical trend observed on the 176 data sets. This gain is partially mitigated on the statistical analysis owing to the fact that only 25\% of the data were noticeably contaminated by the LWE. Moreover, the atmospheric part of the PSF model can mimic PSF asymmetries by modifying the ratio $\alpha_x/\alpha_y$ in Eq. \ref{E:psd}. Despite it leads to an inaccurate representation of the LWE impact on the PSF, it permits to mitigate the SR and FWHM estimation error comparatively to a symmetric atmospheric PSF model. As a summary, fitting the twelve extra parameters to represent the LWE allows can enhance the PSF metrics estimation by a factor up to five.}

\begin{table*}[h]
    \centering
    \begin{tabular}{||c||cc||cc||cc||c||}
     \hline
    \hline
    & \multicolumn{2}{c||}{Case 1} & \multicolumn{2}{c||}{Case 2} & \multicolumn{2}{c||}{Case 3} &  \makecell[bb]{ \multirow{4}{*}{ \includegraphics[width=0.047\linewidth]{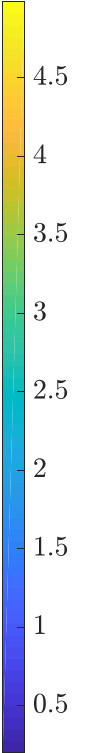}} \vspace{-0.65cm}} \\
    \hline
    \hline
    
    \makecell[bc]{Sky \\ image \vspace{1cm}} & \multicolumn{2}{c||}{\includegraphics[width=0.2\linewidth]{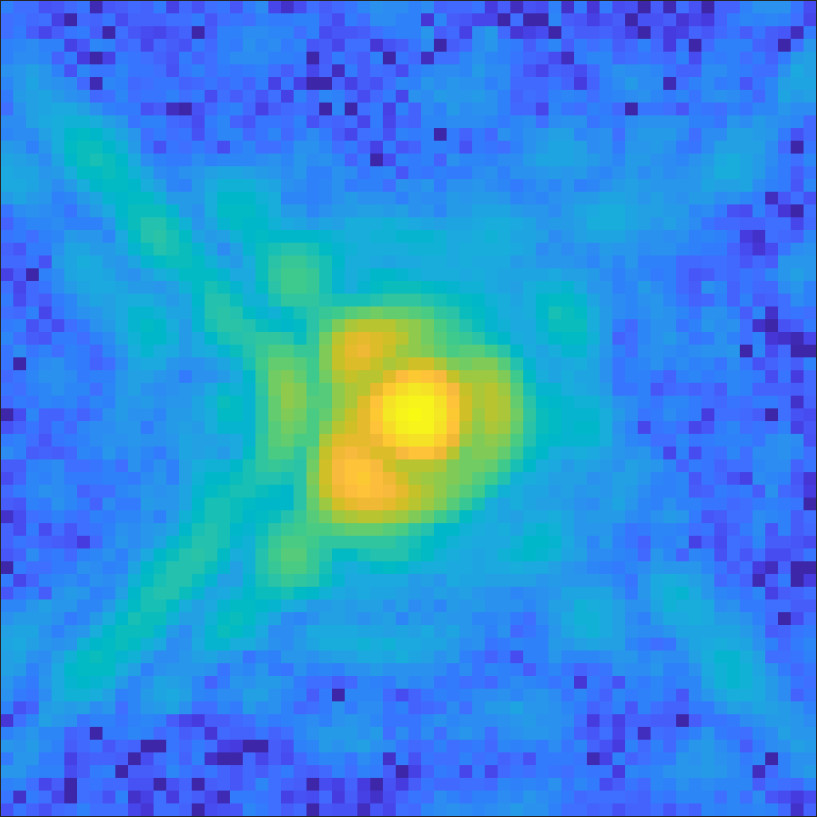}} & \multicolumn{2}{c||}{\includegraphics[width=0.2\linewidth]{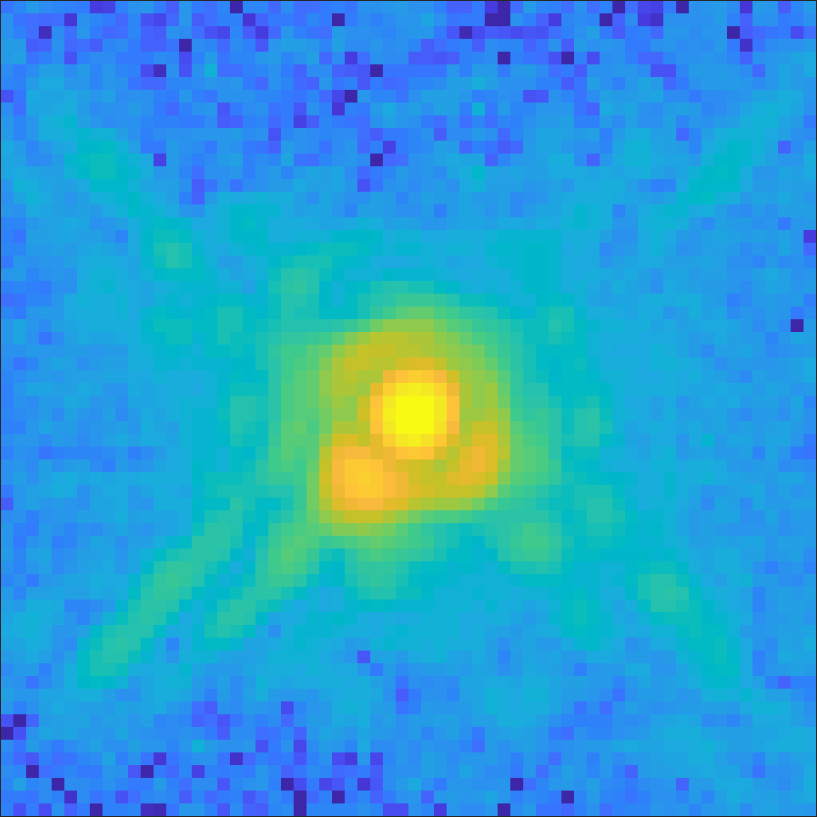}}  & \multicolumn{2}{c||}{\includegraphics[width=0.2\linewidth]{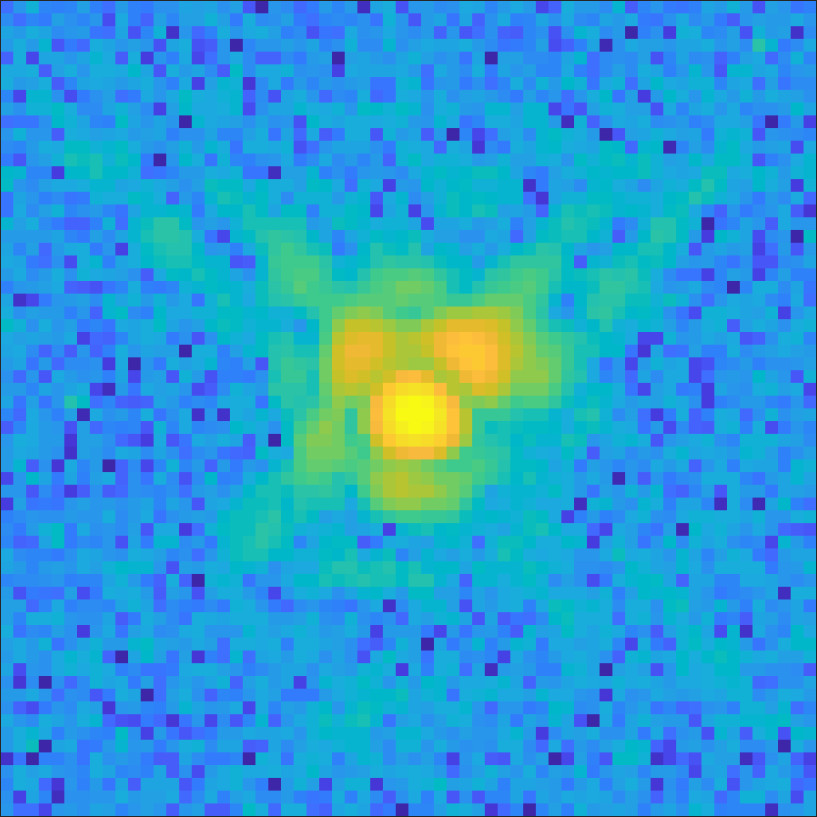}} & \\

    \makecell[bc]{Model \vspace{.8cm}}& \includegraphics[width=0.1\linewidth]{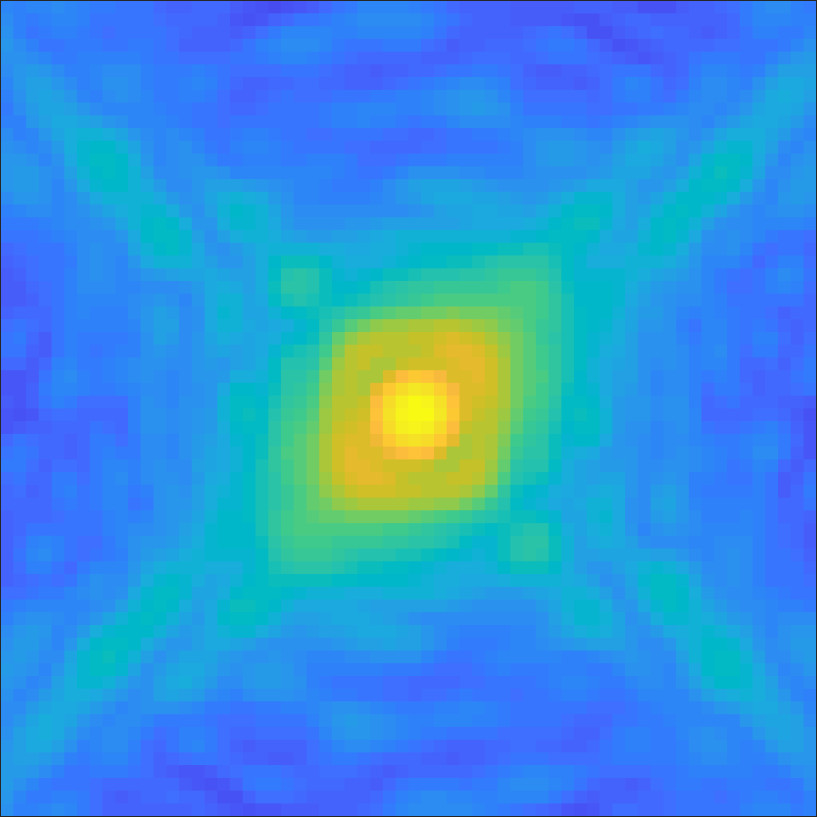} & \includegraphics[width=0.1\linewidth]{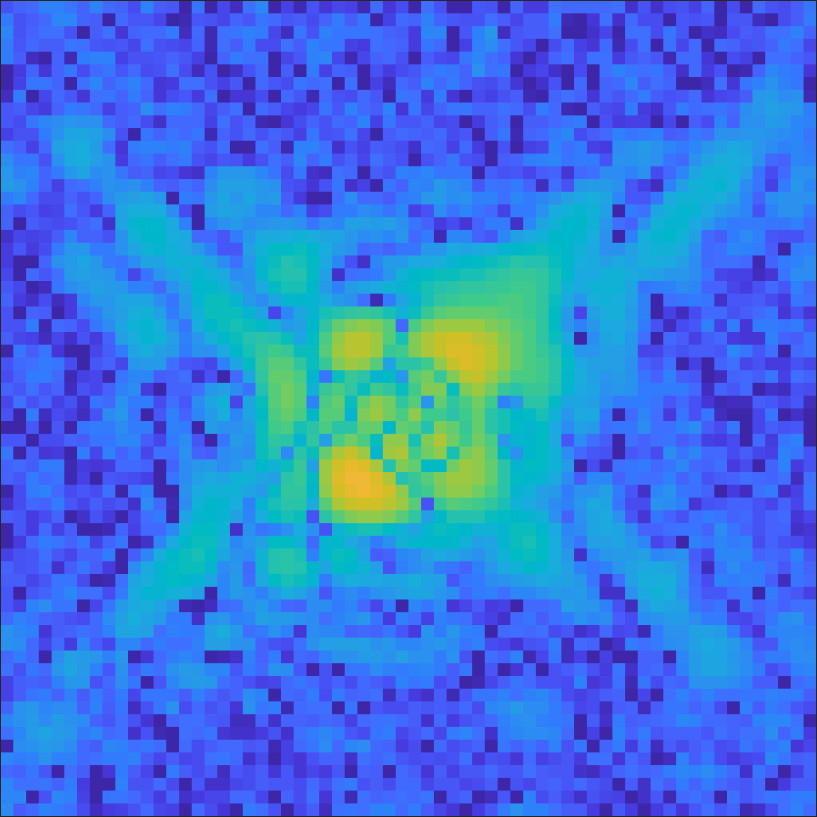}&\includegraphics[width=0.1\linewidth]{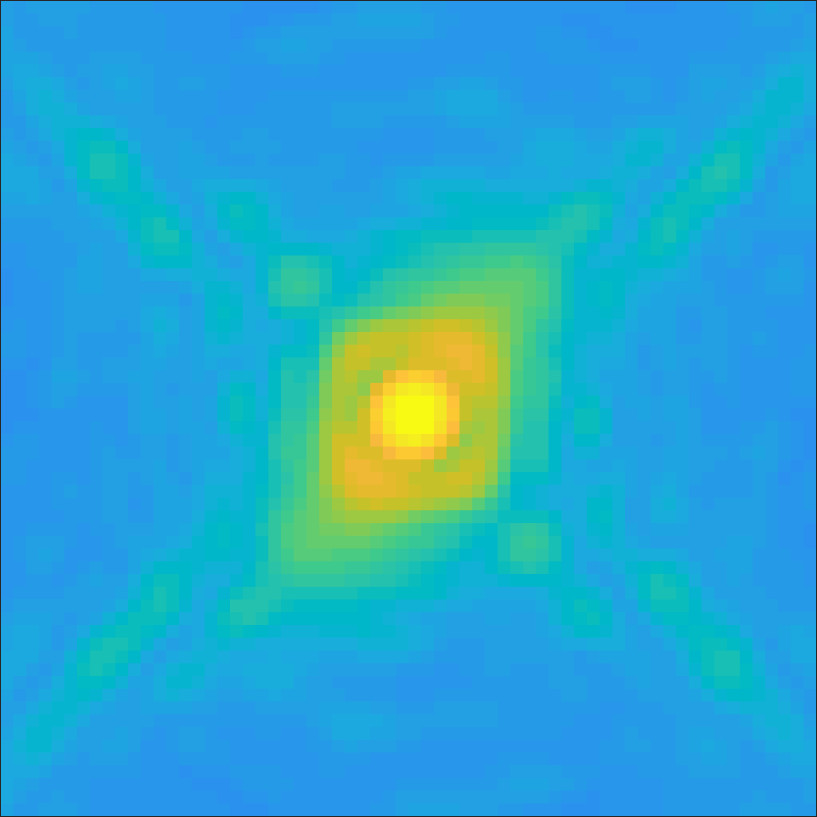}& \includegraphics[width=0.1\linewidth]{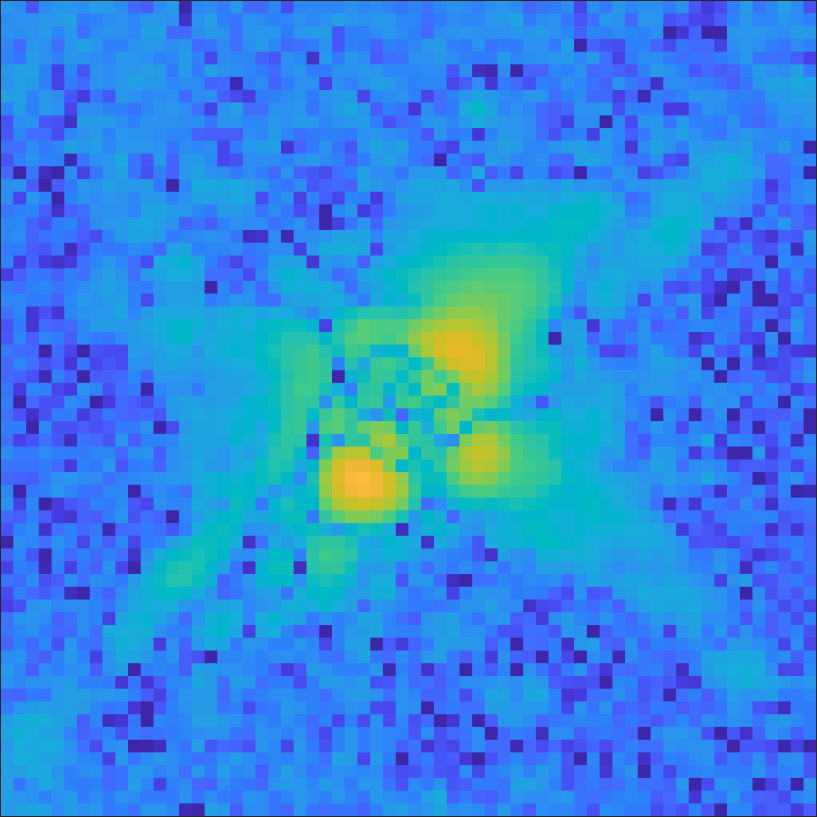}& \includegraphics[width=0.1\linewidth]{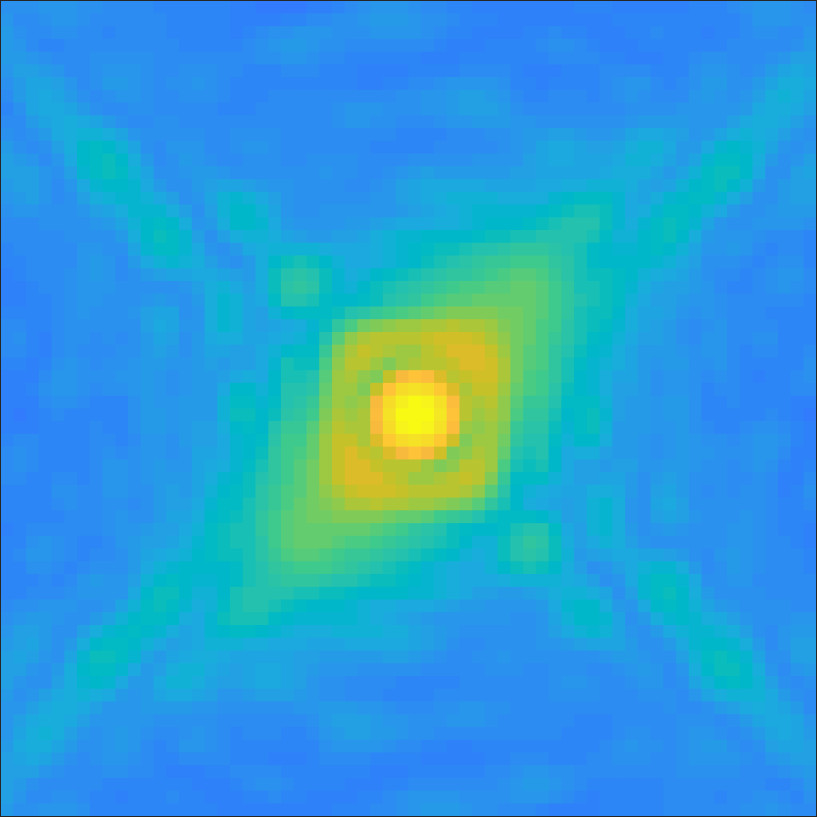}& \includegraphics[width=0.1\linewidth]{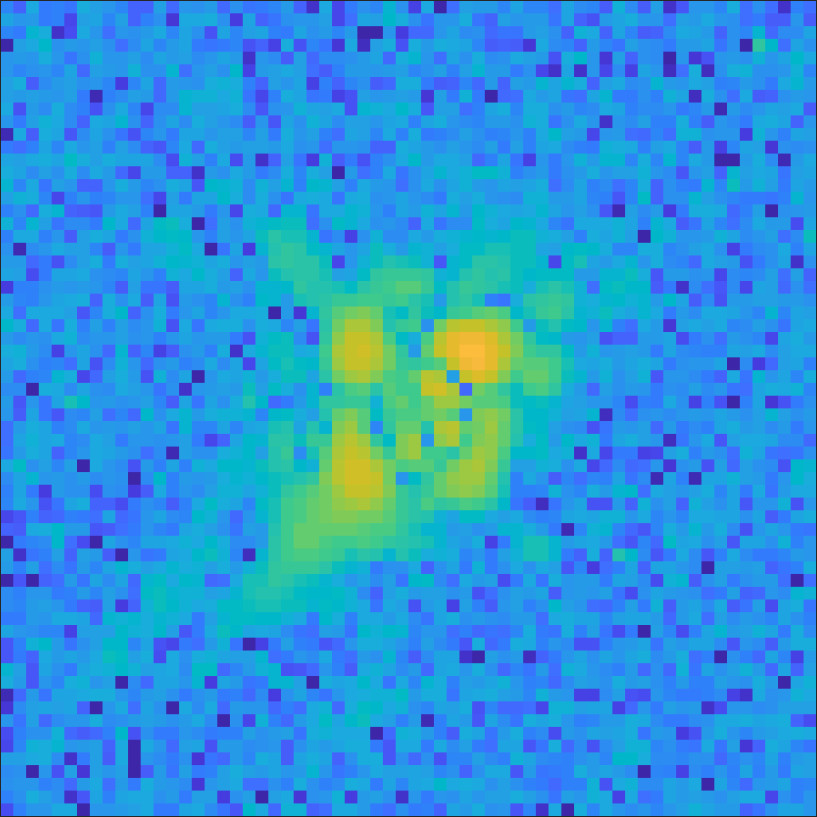} &\\

    \makecell[bc]{Model \\ +\\ LWE \vspace{.3cm}}& \includegraphics[width=0.1\linewidth]{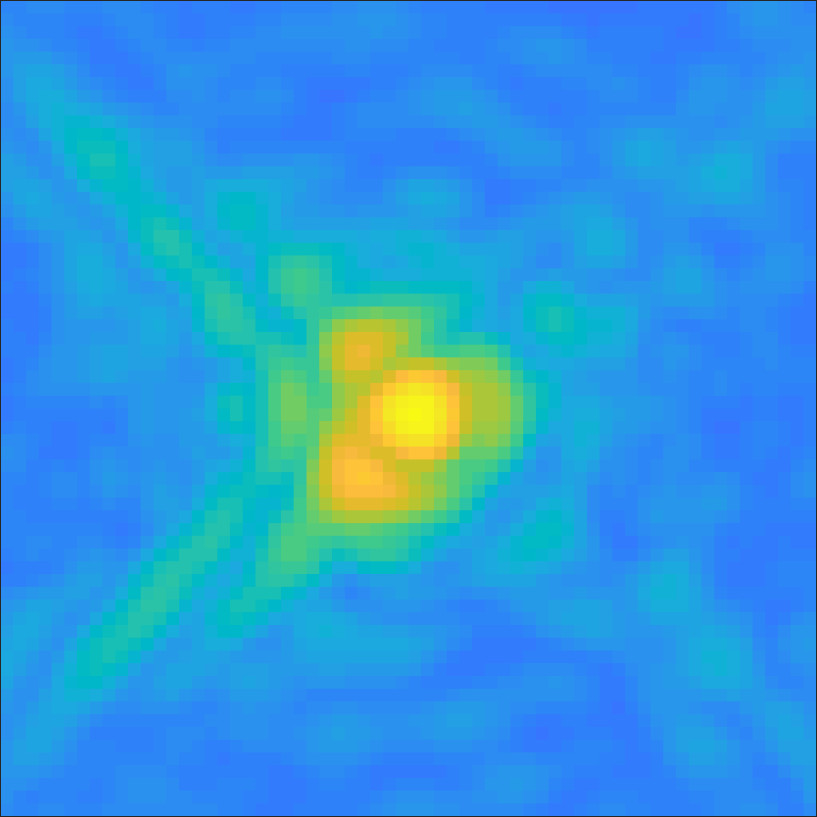} & \includegraphics[width=0.1\linewidth]{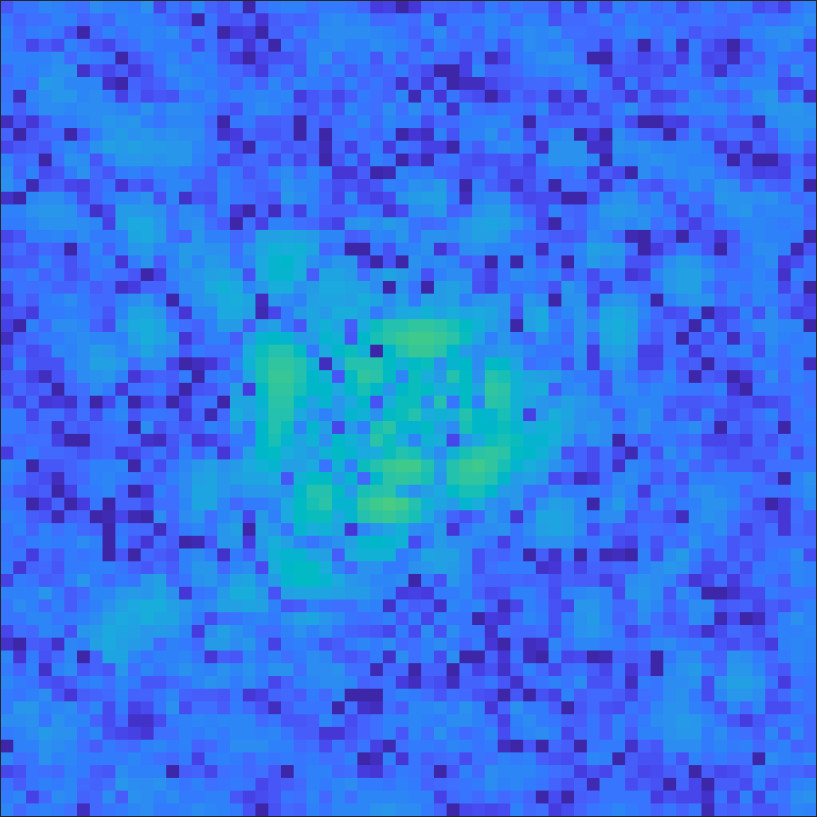}&\includegraphics[width=0.1\linewidth]{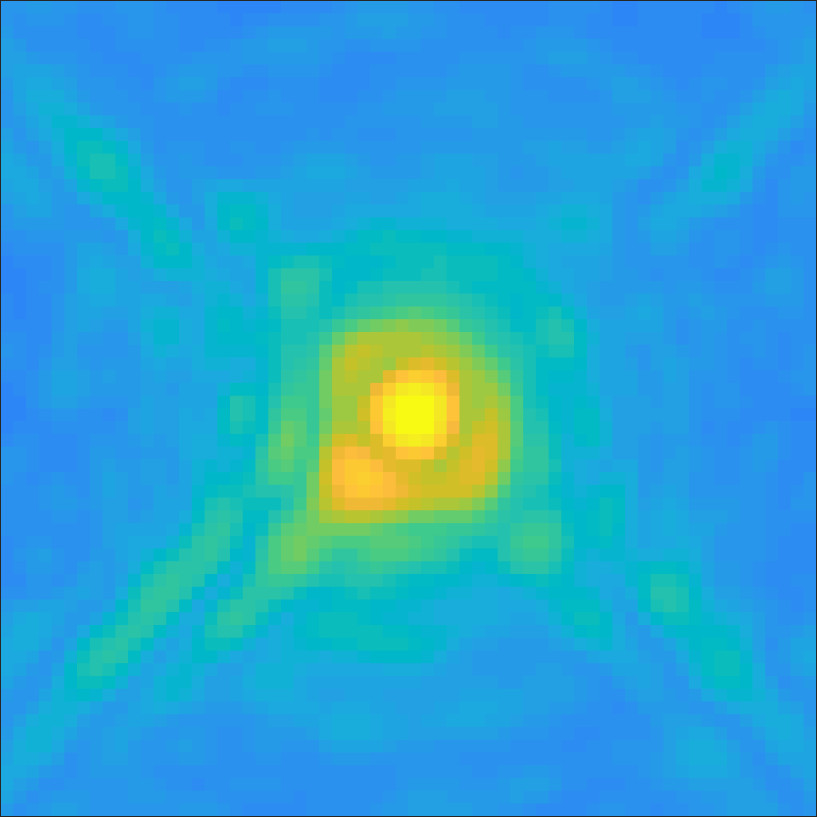}& \includegraphics[width=0.1\linewidth]{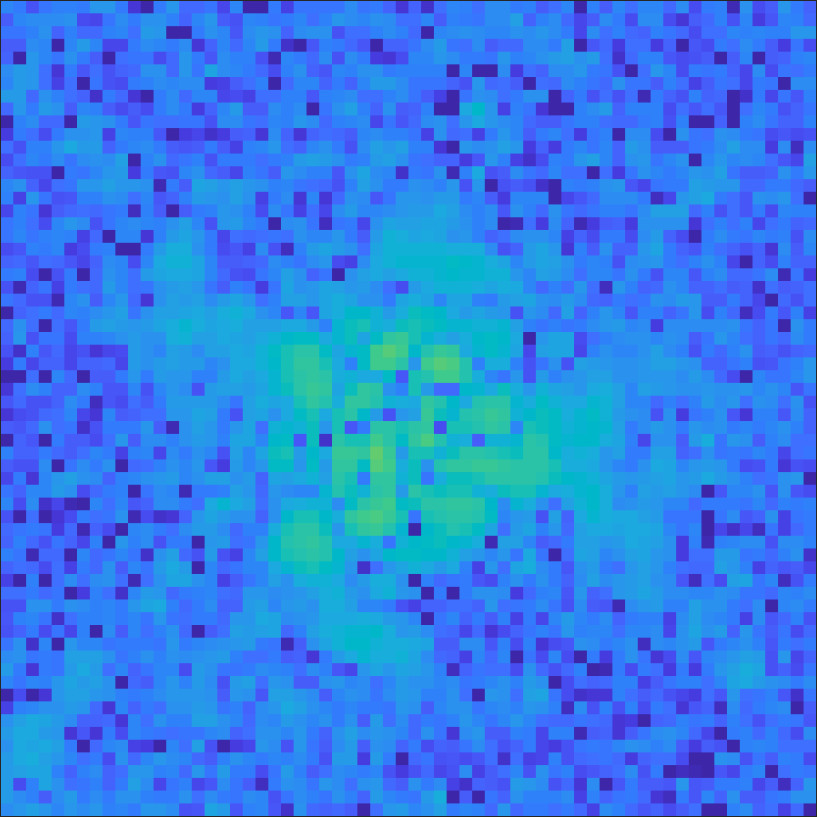}& \includegraphics[width=0.1\linewidth]{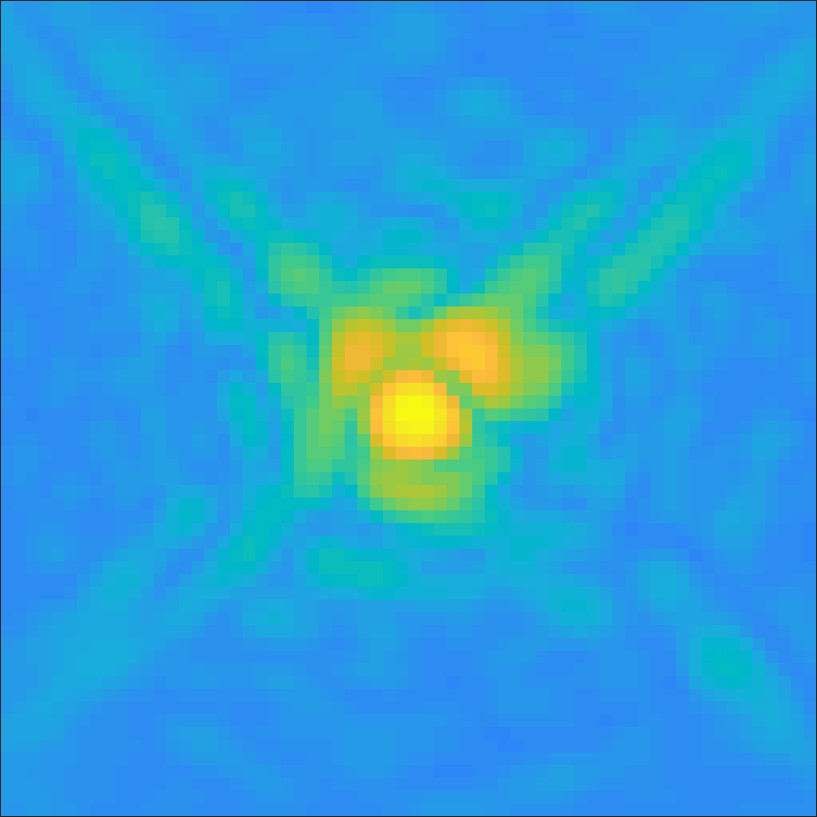}& \includegraphics[width=0.1\linewidth]{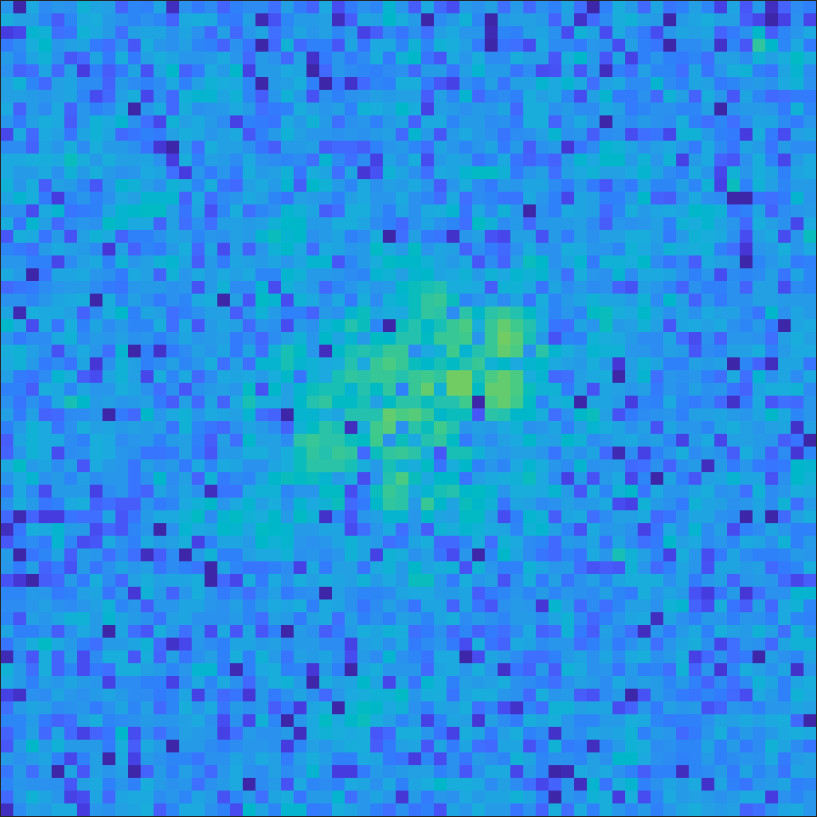} & \\
    \hline
    \hline
    \end{tabular}
    \captionof{figure}{Comparison in log scale of on-sky PSFs and fitted model considering the LWE or not. Each column correspond to a single data sample acquired at different epoch. For each model configuration, we report both the adjusted PSF and the residual map. This figure illustrates that the LWE retrieval, modelled by a differential piston, tip and tilt improves drastically and consistently the residual map.}
    \label{fig:psf_lwe}
\end{table*}

\begin{table}[h]
    \centering
    \begin{tabular}{|c|c|c||c|c||c|}
    \hline
    Strategy &\multicolumn{2}{c||}{ $\Delta$ SR (pts)} & \multicolumn{2}{c||}{ $\Delta$ FWHM (mas)}  & MSE\\
    \hline
    (SR>40\%)& Median & std & Median & std &     \\ 
    \hline
    No LWE & 0.5 & 1.9 & -0.2 & 1.2 & 1.5\\
    \hline
    With LWE & 0.2 & 1.0 & -0.1 & 1.0 & 0.7\\
    \hline
    \end{tabular}
    \caption{Median and 1-$\sigma$ dispersion of SR and FWHM errors obtained over the 176 H-band and high SR (>40\%) SPHERE/IRDIS images treated in this analysis for the two implemented strategies: (i) PSD/stellar parameters retrieval only (11 parameters in total) and (ii) including piston, tip, and tilt retrieval for the four pupil segments (23 parameters in total). Systematically, accounting for the LWE model allows us to reduce the biases and the dispersion on PSF estimates.}
    \label{tab:srsphere}
\end{table}

\begin{table}[h]
\ccc{
    \centering
    \begin{tabular}{|c|c|c|c|c|}
    \hline
    Case & LWE & $\Delta$ SR (pts) & $\Delta$ FWHM (mas) & MSE (\%) \\
    \hline
    \multirow{2}{*}{1} & No & -1.3 & 1.5 & 1.7\\
    & Yes & -0.3 & 0.3 & 0.2 \\
    \hline
     \hline
    \multirow{2}{*}{2} & No &  0.1 & 0.4&  1.6\\
    & Yes &  0.04& 0.1&0.3 \\
    \hline
     \hline
    \multirow{2}{*}{3} & No & 0.8&  -1.5& 2.1\\
    & Yes & 0.6& -1.4 &0.4\\
    \hline
    \end{tabular}
    \caption{Impact of the LWE fit compared to a pure atmospheric model on SR, FWHM and MSE metrics for the three cases presented in Fig. \ref{fig:psf_lwe}.}
    \label{tab:case_lwe}}
\end{table}

\begin{figure*}[h]
    \centering
    \includegraphics[width=\linewidth]{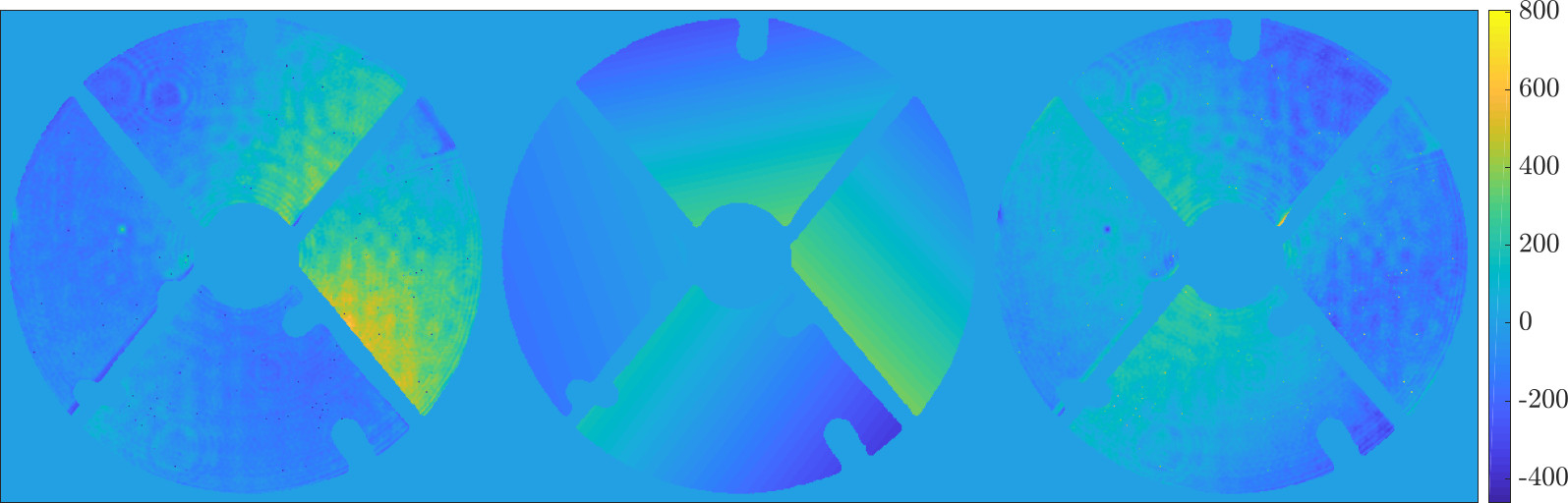}
    \hspace{.2cm}
    \includegraphics[width=\linewidth]{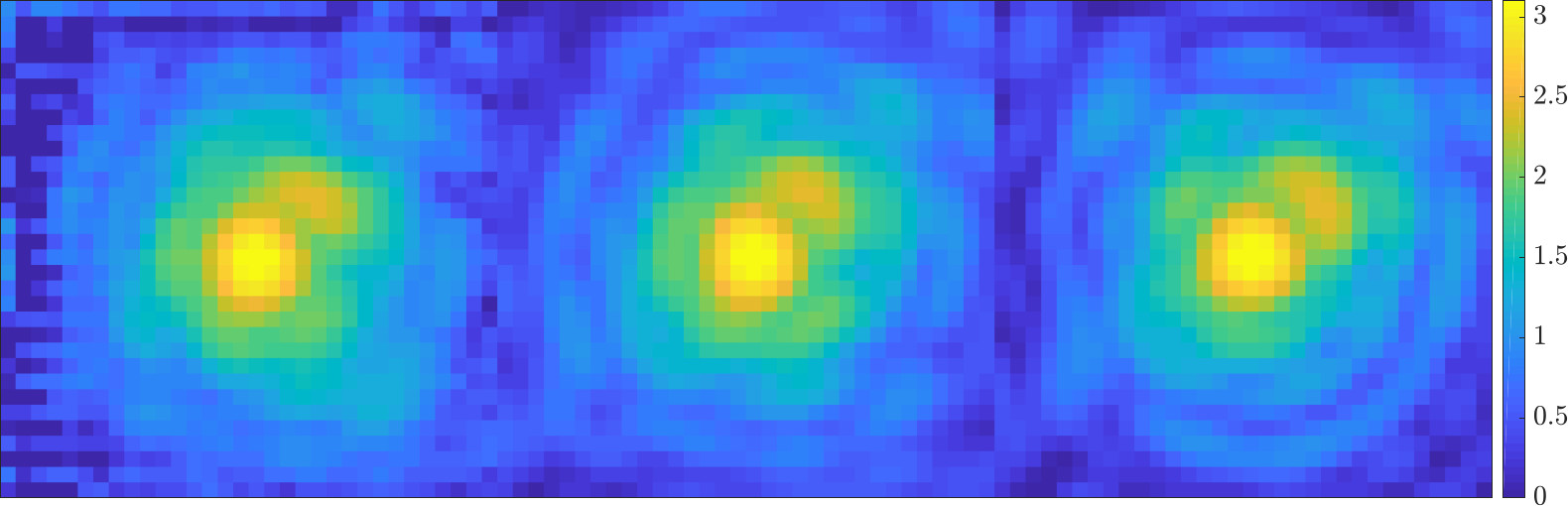}
    
    \caption{\textbf{Top:} From left to right: ZELDA measurements (114\,nm std), retrieved static map (140\,nm) from the PSF-fitting and residual (70\,nm std) in nm. This comparison emphasizes the meaningfulness of the PSF-fitting outputs, which compares very well with dedicated measurements of the LWE. \textbf{Bottom:} From left to right: On-sky DTTS image, best-fitted PSF, ZELDA-based PSF.}
    \label{fig:lwemap}
\end{figure*}

In order to confirm the robustness of the LWE retrieval, we have compared the estimated static aberration map with ZELDA measurements \citep{Vigan2019} taken during SPHERE commissioning nights in 2014. The PSF-fitting was performed using \ccc{the differential tip-tilt sensor (DTTS) that delivers 32$\times$32 pixels images \citep{Baudoz2010,Sauvage2015} acquired simultaneously with SPHERE/IRDIS using the ZELDA focal-plane mask to measure optical aberrations within the pupil-plane}. Moreover, in order to calibrate ZELDA measurements (that are in ADU) to reconstruct the phase, we followed the process described by \citet{Sauvage2015} and removed the mean ADU value over the pupil and adjusted a multiplicative factor to obtain the closest PSF possible from the DTTS observation as reported in Fig. \ref{fig:lwemap}. According to this calibration, we obtained standard deviations of 173\,nm rms and 146\,nm rms on the ZELDA map and the DTTS image-based map,respectively, which leads to a quadratic difference of 92\,nm rms. This residual includes the internal aberrations in the IRDIS science path that do not impact the DTTS images and reach 50\,nm and standard deviation\citep{Beuzit2019}. \ccc{Also, this residual is likely mostly due to higher order aberrations not included in the differential piston, tip, and tilt of the static map model, which suggests that further improvements can be pursued to characterize the LWE.} 
In conclusion, the framework presented in this paper is a powerful tool to assist in the estimation of internal aberrations. \ccc{Furthermore, we are able to assess these aberrations from the DTTS image that delivers a post-AO and non-coronagraphic image regardless the coronagraphic mask employed during SPHERE/IRDIS observations. Consequently, this technique allows the joint estimation of atmospheric and instrumental defects using tip-tilt sensors measurements. Future work will address the extension of this strategy to ELT instruments that will rely on a 2x2 Shack-Hartmann WFS to measure low-order modes and that will provide PSFs from which we will be able to assess the telescope aberrations and calibrate a PSF model for science exploitation.}

\section{Conclusions}

This paper revisits and improves the analytical framework proposed by \citep{Fetick2019_Moffat_aa}, which now includes a parametrization of static aberrations for a joint retrieval of atmospheric parameters, AO performance and static aberrations.

We demonstrate in an exhaustive manner, using 4812 PSFs obtained from four different observatories and seven optical or NIR instruments, that the proposed model matches the PSF of any AO flavor within 4\% error, even for high-SR observations. We also illustrate that the retrieved parameters carry relevant information about the AO performance and the atmospheric conditions, especially seeing and wavefront error, that shows agreement with the literature.

Finally, we illustrate that this model, upgraded with additional degrees of freedom to estimate static aberrations, allows the atmospheric parameters, AO performance and static aberrations, to be retrieved simultaneously. Particularly, the framework presented in this paper allows us to assess (i) the Keck pupil segment piston errors and especially the presence of a stair mode that was already pointed out \citep{Ragland2018_COPHASING} and, (ii) the LWE on SPHERE/IRDIS images as a combination of differential piston, tip and tilt over the four VLT pupil segments delimited by the spiders.

This model is a unique tool that gathers an AO diagnosis and a PSF estimation facility in the simplest and the most parsimonious way possible. However, we have handled it as a parametric model so far and the next step of this work is to enable a forward estimation of its parameters from contextual data expect the focal-plane image. We emphasized that the connection of these parameters with the observing conditions is not easily made; nevertheless, thanks to the parsimony of this model, this regression problem consists in assessing a few parameters (up to a few tens) from a large amount of data, which is provided by the AO telemetry, all the sensors within the telescope and the dome and the external meteorological profilers, which can be achieved with the use of neural networks. We are currently developing convolutional neural networks capable of directly estimating the model outputs from either an imaged PSF or a subsample of AO telemetry and we will present this work in a dedicated publication.

\begin{acknowledgements}
      This work has been partially funded by the French National Research Agency (ANR) program APPLY - ANR-19-CE31-0011. This work also benefited from the support of the WOLF project ANR-18-CE31-0018 of the  and the OPTICON H2020 (2017-2020) Work Package 1. This work has made use of the SPHERE Data Centre, jointly operated by OSUG/IPAG (Grenoble), PYTHEAS/LAM/CeSAM (Marseille), OCA/Lagrange (Nice) and Observatoire de Paris/LESIA (Paris) and supported by a grant from Labex OSUG@2020 (Investissements d’avenir – ANR10 LABX56).
      Authors thank G. Fiorentino (INAF), F. Kerber (ESO), D. Massari (U. Bologna, INAF, Kapteyn Astronomical Institute), J. Milli (IPAG), and E. Tolstoy (Kapteyn Astronomical Institute) to have conducted the SPHERE/ZIMPOL observations of NGC 6121 in 2018 and provided the data. Authors thank P. Vernazza (LAM) to have delivered SPHERE/ZIMPOL PSF calibrators obtained during his large programme. Authors acknowledge the contribution of J. Milli (IPAG) to utilize the SPHERE Data centre pipeline. Authors thank F. Cantalloube (MPIA) and M. N'Diaye (Lagrange) to provide the APLC mask model.  Authors thank Jean-François Sauvage (ONERA/LAM) for providing SPHERE DTTS and ZELDA measurements that served in the Sect. \ref{SS:lwe}. Authors thank Sam Ragland (W.M. Keck Observatory) for giving access to NIRC2 images obtained during PSF reconstruction engineering nights on Keck II. Author thank G. Agapito (INAF) and E. Pinna (INAF) to communicate SOUL/LUCI commissioning data. Authors thank J. Vernet (ESO) and S. Oberti (ESO) for providing MUSE NFM commissioning data. Authors thank M. Andersen (Gemini), G. Sivo (Gemini) and A. Shugart (Gemini) for communicating GeMS/GSAOI observations of Trumpler 14 and supporting the data processing. Author express their gratitude to F. Pedreros-Bustos (LAM) for a fruitful revision of the manuscript.
\end{acknowledgements}

\bibliographystyle{aa}
\bibliography{biblioLolo}
\end{document}